\documentclass[twocolumn, twocolappendix]{aastex702}
\usepackage[utf8]{inputenc}
\usepackage{amsmath}	% Advanced maths commands
\usepackage{amssymb}
\usepackage{enumitem}

\graphicspath{{./}{figures/}}

\begin{document}

\title{Marginalized planet-to-star radius ratio posteriors for low-SNR transiting planets via dimensionality reduction}

%% The \author command is the same as before except it now takes an optional
%% argument which is the 16 digit ORCID. The syntax is:
%% \author[xxxx-xxxx-xxxx-xxxx]{Author Name}

\author[0000-0002-2805-9405]{Oryna Ivashtenko}
\affiliation{Department of Particle Physics and Astrophysics, Weizmann Institute of Science, Rehovot 7610001, Israel}
\email{oryna.ivashtenko@weizmann.ac.il}

\author{Barak Zackay}
\affiliation{Department of Particle Physics and Astrophysics, Weizmann Institute of Science, Rehovot 7610001, Israel}
%\collaboration{}{()}
\email{barak.zackay@weizmann.ac.il}

\begin{abstract}
Occurrence rate estimates of small long-period planets rely on the radii of low signal-to-noise ratio transit candidates whose posteriors should be propagated into the population inference. These posteriors should include the error budget from all latent parameters and account for the effects of the correlated noise. Full Markov chain Monte Carlo fits with Gaussian processes take days of runtime per target, and provide information on latent parameters that will be marginalized over anyway. We present a method that computes the marginalized posterior of the planet-to-star radius ratio while sampling only a lower-dimensional effective parameter space. 
%The transit depth is treated semi-analytically using the approximate linearity of the transit model, the transit epoch is marginalized using the convolution theorem, and the orbital and stellar parameters are compressed into a template bank parametrized by the transit duration and impact parameter, with priors that preserve the probability mass of the original parameters. 
The 11-dimensional problem reduces to a 4-dimensional one, while the posterior retains the uncertainty budget of all the nuisance parameters, including eccentric orbits, stellar parameters, and limb darkening. The method operates on whitened light curves, mitigating biases due to correlated noise. We show that the posteriors are calibrated on simulated systems: the true radii have uniformly distributed p-values. Applied to 419 faint long-period Kepler Objects of Interest, the method yields radius ratios broadly consistent with the KOI table, although their differences exceed those expected from the reported uncertainties.
\end{abstract}


\section{Introduction} 
\label{sec:intro}

\paragraph{Context of the problem: occurrence rates}
Estimating the exoplanet occurrence rates using the Kepler mission \citep[][]{borucki_2003_kepler_mission} data remains a topic of ongoing research \citep[][]{kunimoto_2020_occurrence_fgk, dattilo_2023_occurrence, robnik_2025_nst}. Kepler is still a unique mission for statistical studies of small long-period planets due to its unprecedented photometric precision, large uniform stellar sample, long continuous observation time and well-characterized detection sensitivity. In addition, development of statistical methods with Kepler data serves as a preparation for the analysis of the future missions such as PLATO \citep[][]{margin_2018_plato} and Earth 2.0 \citep[][]{ge_2024_earth_2.0}, aiming to uncover a new population of rocky planets.

An accurate and precise occurrence rate measurement requires several components:
\begin{itemize}[noitemsep, topsep=0pt, left=1pt, label=-]
    \item A homogeneous search and vetting over the target catalog;
    \item Reliability and completeness products for search and vetting;
    \item Precise stellar parameters;
    \item Precise planetary parameters;
    \item Good selection of occurrence rate estimation method and model;
    \item Target selection characterization.
\end{itemize}
Each component of this chain is being continuously improved, with new detection and vetting pipelines developed \citep[][]{robnik_2026_detection_vetting_pipeline, ivashtenko_2025}, new statistical approaches to occurrence rates \citep[][]{hsu_2018_occurrence_bayesian, bryson_2020_occurrence_probabilistic}, investigations of the reliability effects \citep[][]{bryson_2020_reliability}, precise measurements of stellar properties \citep[][]{berger_2020}. 

For small long-period planets, the occurrence rate estimation is especially challenging due to the low number of detections, low signal-to-noise (SNR) complicating the vetting and the parameter estimation, and low reliability and completeness in this regime \citep[][]{bryson_2020_reliability, bryson_2020_occurrence_probabilistic, thompson_2018}. In \citet[][]{ivashtenko_2025}, we developed a search pipeline with controlled reliability to reduce the false alarm contamination of the small-planet catalog. This pipeline also uses injection-recovery and an empirical background distribution to estimate, for each candidate, the probability of being a genuine planet rather than a noise-induced detection. We intend to use the resulting candidate catalog and these probabilities to infer planet occurrence rates, allowing even low-SNR candidates that cannot be individually confirmed to contribute probabilistically to the measurement. Such an occurrence-rate analysis requires posterior distributions for the radii of the candidate planets. In this paper, we present an efficient and robust method for obtaining these radius posteriors.

\subsection{The role of radius posteriors in occurrence rate inference}
\label{sec:posteriors_significance}

The primary occurrence rate product for transiting planets is the occurrence rate as a function of orbital periods and planetary radii \citep[e.g.][]{bryson_2021_occurrence_kepler, dattilo_2023_occurrence}. This choice is natural, since the period and radius are the parameters most directly constrained by the transit periodicity and depth \citep[e.g.][]{seager_2003_transit_parameters}. The orbital periods are usually measured precisely, apart from possible errors when a harmonic of the true period is found \citep[e.g.][]{petigura_2013_plateau}. The error in the period measurement \citep[e.g. the error reported in the NASA Exoplanet Archive][]{akeson_2013_nasa_archive} is typically much smaller than the characteristic period bins used for occurrence rate measurements \citep[e.g.][]{hsu_2018_occurrence_bayesian, kunimoto_2020_occurrence_fgk, dattilo_2023_occurrence}. 

The error in the planetary radius can exceed 10\% \citep[e.g.][]{bryson_2020_occurrence_probabilistic} and may lead to misattribution of planets to incorrect bins, if binning is used \citep[][]{kunimoto_2020_occurrence_fgk}. In \citet[][]{bryson_2020_occurrence_probabilistic} and \citet[][]{foreman_mackey_2014_population_inference}, it was shown that neglecting the radius uncertainty can bias the occurrence rate and underestimate its dispersion. 

Historically, in the planetary population inference, planetary radii were treated as point estimates (maximum a posteriori or median values) \citep[e.g.][]{zhu_dong_2021, petigura_2013_plateau}. Modern studies propagate the planetary radius uncertainty, for example the
Approximate Bayesian Computation (ABC) framework \citep[][]{hsu_2018_occurrence_bayesian, kunimoto_2020_occurrence_fgk}. In it, the forward model simulates true underlying planets, assigns them noisy observed radii sampled from analytic uncertainty prescriptions, and bins them according to these radii. 

Using the full posteriors, as prescribed by \citet[][]{foreman_mackey_2014_population_inference}, is still not common in planetary population inference, but is gaining acceptance~\citep[e.g.][]{van_zandt_2026_occurr}. In other fields, this approach is standard, for example, in gravitational wave astronomy \citep[][]{mandel_2019_gw_distributions, lvk_2026_gwtc5_population}. In this prescription, the hierarchical Bayesian inference framework models the population as a Poisson process and derives the rates by marginalizing over the posterior distribution of each detection. Effectively, it makes every candidate contribute to several parameter space bins according to the weight of the posterior, and avoids the artifacts of assigning one point to one bin described in \citet[][]{hsu_2018_occurrence_bayesian}.
We intend to use this approach in our future work to incorporate the catalog information into the population inference more efficiently and avoid biases and artifacts. 

Thus, we need uniformly measured planetary radius posteriors for all candidates in our catalog. The posterior width should incorporate the entire error budget that affects the planetary radius measurement. 
This is particularly important in regions of parameter space with low completeness, where only a few detections shape the overall occurrence rate, and posterior tails can contribute substantial probability mass.

\subsection{Motivation for a reduced-dimensional inference}
\label{sec:motivation}

\paragraph{The inference problem}
Multiple degenerate parameters influence the measured transit properties: the stellar mass, radius, effective temperature, surface gravity, and metallicity; the orbital period, transit epoch, inclination, eccentricity, and periastron argument; the planetary radius; and other parameters in more complex models. Most stellar parameters influence the model only by means of the effective stellar density and the limb darkening coefficients \citep[][]{li_2019_dv_fit, rowe_2015_koi_mcmc, lissauer_2024}. The limb darkening coefficients are degenerate with the impact parameter in how they influence the transit shape, and the impact parameter is degenerate with the planet radius in how they influence the transit depth. Transits offer very limited information about orbital elements, such as the eccentricity and periastron argument \citep[][]{dawson_2012_photoeccentric_effect}. In the low-SNR regime, the transit shape is poorly resolved, the data constrain almost no physical parameters, and the influence of priors is significant.

In addition, Kepler light curves contain correlated (or red) noise \citep[][]{jenkins_2002, jenkins_2010_pipeline} that impacts both the detection and the parameter estimation \citep[][]{li_2019_dv_fit, matesic_2024}. Approaches to treat it include Gaussian processes \citep[][]{matesic_2024, almenara_2026}, pre-whitening \citep[][]{li_2019_dv_fit}, wavelet analysis \citep[][]{carter_2009, csizmadia_2020}, empirical noise variance correction \citep[][]{wong_2020}, and detrending followed by a white-noise likelihood \citep[][]{rowe_2014_koi_validation_and_fit, alexoudi_2022, niraula_2017, van_eylen_2015, van_eylen_2019}. Detrending with a sliding average or polynomials does not properly account for the frequency structure of the noise and leaves residual low-frequency noise. Gaussian processes require a proper model choice to avoid overfitting and are preferably fit simultaneously with the other parameters \citep[][]{aigrain_2023_gp, barros_2020}, which increases the fit complexity.

\paragraph{Current practice}
For individual systems, full posterior evaluations are routinely performed \citep[e.g.][]{almenara_2026} with MCMC or nested sampling codes \citep[e.g.][]{foreman_mackey_2013_emcee, gazak_2012_tap, eastman_2013_exofast, eastman_2019_exofast_2, espinoza_2019_juliet, foreman_mackey_2021_exoplanet}. 
% An MCMC fit typically requires tens of walkers with tens of thousands of iterations each, i.e. of order $10^6$ chain elements \citep[][]{wong_2020, alexoudi_2022}, each requiring a transit model generated with a code such as \texttt{batman} \citep[][]{kreidberg_2015_batman} and a likelihood evaluation.
The reported run times depend on the implementation and the hardware, but their dependence on the problem is instructive.
% SRC: Eastman 2013, abstract ("well-mixed in under 5 minutes on a standard desktop computer"); Sect. 3 lists the stepping parameters (log P, log K, sqrt(e)cos w, sqrt(e)sin w, T_C, gamma, gamma-dot, F0, cos i, log(a/R*), p, log g, Teff, [Fe/H]); LD interpolated from Claret tables.
% SRC: Eastman 2019, Sect. 21: 369,356 steps in 4.75 min; Sect. 21: no gain from multiple cores.
A multi-parameter fit with a white-noise likelihood takes on the order of minutes on a typical computer \citep[][]{eastman_2013_exofast, eastman_2019_exofast_2, espinoza_2019_juliet}.
% SRC: Espinoza 2019, Sect. 4 (runtime paragraph, p. 2280): "for single photometry and/or radial-velocity analyses juliet takes of order of minutes"; K2-140b "took several hours"; "problems with dimensions of order ~20 take several hours ... if several GPs are included ... of order a day"; Table 1 lists the photometric parameters.
% A nested-sampling white-noise likelihood fit of one photometric data set for the period, epoch, radius ratio, impact parameter, stellar density, two limb darkening coefficients, and three instrumental parameters takes of order minutes on a laptop; 
Problems with $\sim20$ parameters and several Gaussian processes take up to a day \citep[][]{espinoza_2019_juliet}.
% SRC: Eastman 2019, Sect. 26 (40 min vs just under 2 h for HAT-P-3b global fit depending on starting values); Sect. 23.3 ("weeks", K2-266 60 days); Sect. 27 (TOI fits: 2.5-day limit, "most fits reach the 2.5 day runtime limit", "Some 5-10% of fits fail catastrophically").
Global fits that add a spectral energy distribution and stellar evolutionary tracks take from tens of minutes to hours for a single planet, depending on the initialization, weeks for complex multi-planet systems, and the automated fits of all TESS Objects of Interest are capped at 2.5 days per target, a limit that most of them reach before converging \citep[][]{eastman_2019_exofast_2}.
% SRC: Matesic 2024, Table 1 / Sect. 2.2 (rho*, q1, q2, T0, P, b, Rp/R*, F0, sigma_w, sigma_c, l_c; circular orbits assumed); Sect. 6: "When allocated one node (32 CPU threads) ... typical per-target timescales on the order of a week".
For low-SNR long-period Kepler candidates, a fit of seven transit parameters on a circular orbit with a three-parameter Gaussian process noise model, sampled with nested sampling to obtain the Bayesian evidence for the planet hypothesis, takes of order a week per target on a 32-thread node \citep[][]{matesic_2024}.
% SRC: Ford 2005 Sect. 3.2 ("Poor choices can lead to extremely inefficient sampling and hence slow convergence"); Eastman 2013 Sect. 4.1 and Table 1 (steps to convergence vs eccentricity parametrization); MacDougall 2023 Sect. 3 ("convergence issues due to the high curvature of the posterior parameter space").
The cost thus grows with the number of parameters and with the complexity of the noise model, while the convergence of the samplers degrades with the correlations and the curvature of the posterior \citep[][]{ford_2005, eastman_2013_exofast, mac_dougal_2023}.

\paragraph{Survey catalogs}
Survey catalogs use reduced models, for example, by assuming a circular orbit, fixing stellar parameters, or using directly measurable values such as transit duration, ingress/egress times, or impact parameter. The uniform Kepler analysis assumed circular orbits and fixed the limb darkening coefficients to tabulated values, sampling six parameters with $\sim10^6$ chain elements per candidate \citep[][]{rowe_2014_koi_validation_and_fit, rowe_2015_koi_catalog, coughlin_2016_dr24, thompson_2018}.
% SRC: Hoffman & Rowe 2017 (KSCI-19113-001): "KOIs are not modeled with MCMC analysis when the transit event does not have sufficient S/N for proper modeling (S/N >~ 7) ... or the Markov-Chain did not converge"; Rowe 2015 Sect. 5.1; Lissauer 2024 Sect. 2.3: "There are 220 KOIs flagged as 'P' or 'S' without MCMC computed posteriors"; Li 2019 Sect. 9: 34,032 TCEs; Thompson 2018: 8,054 KOIs.
% It was run on the $\sim8000$ Kepler Objects of Interest (KOIs) rather than on the 34,032 threshold crossing events \citep[][]{li_2019_dv_fit, thompson_2018}, KOIs with $\mathrm{S/N}\lesssim7$ were not modeled, and unconverged chains left candidates without posteriors \citep[][]{rowe_2015_koi_catalog, hoffman_2017_mcmc_notes, lissauer_2024}. 
The Kepler Data Validation pipeline reports best-fit values from a damped least-squares fit with uncertainties from analytically computed Jacobians \citep[][]{li_2019_dv_fit}, and the TESS Objects of Interest catalog reports the pipeline fit values \citep[][]{guerrero_2021_toi}. For the Nancy Grace Roman Space Telescope \citep[][]{nasa_2019_roman}, maximum-likelihood fits are planned for all candidates, with full posteriors as a later product \citep[][]{rowe_2026_sagan_talk}.

These simplified methods are adequate and very helpful in getting approximate best-fit parameters and errors. However, fixing the omitted parameters rather than marginalizing over them underestimates the posterior width and biases the best-fit values \citep[e.g.][]{rowe_2014_koi_validation_and_fit}.
% SRC: Carter & Winn 2009, Sect. 4.2: for equal parts white and 1/f noise, "the white analysis gave an estimate of t_c that differs from the true value by more than 1 sigma nearly 80% of the time"; Gazak 2012: error bars "scaled down by a factor of 0.7" when red noise is set to zero.
% A white-noise analysis of a light curve with equal parts of white and $1/f$ noise places the true transit time outside the $1\sigma$ interval nearly $80\%$ of the time \citep[][]{carter_2009}, and ignoring the red noise shrinks the error bars by a factor of $0.7$ \citep[][]{gazak_2012_tap}. 
A white-noise analysis of a light curve with red noise may significantly misestimate the parameters \citep[][]{carter_2009} and shrink the error bars \citep[][]{carter_2009, gazak_2012_tap}.
Fixing the limb darkening coefficients to theoretical values biases the radius ratio \citep[][]{csizmadia_2013, espinoza_jordan_2015}. If one fixes the eccentricity to zero, one omits from the posterior the option that the same transit depth could have been caused by a higher eccentricity in periastron with a lower impact parameter and a smaller planetary radius; the Kepler validation analysis had to introduce a nuisance parameter scaling the errors of the stellar density measured under this assumption \citep[][]{rowe_2014_koi_validation_and_fit}.
% SRC: Lissauer 2024 Sect. 2.3 (quoted); existing sentence on DV vs KOI table.
The Data Validation radii are not always in agreement with the KOI table values \citep[][]{cumulative_koi_table} obtained with MCMC \citep[][]{koi_dr25_planet_pars_docum}.

\paragraph{Requirements of this work}
Our goal is to obtain planetary radius posteriors for an unbiased occurrence rate estimate of small planets. We need the posterior of the radius only, marginalized over the other parameters, which are not measurable anyway at low SNR, but whose full error budget has to be propagated. We should avoid both biasing the median radii and underestimating the spread of the posteriors, so the red noise has to be included, and the calibration has to be verified. The inference needs to be uniform over the candidates and cheap enough to be rerun: stellar parameters may be updated, and the population inference depends on the priors that may change \citep[][]{hogg_2010, foreman_mackey_2014_population_inference}.
% , and a fit that samples the stellar density with a prior bakes the stellar parameters of the moment into the posterior \citep[][]{MacDougall2023AccurateModeling}.

Neither of the existing options satisfies these requirements. Reduced fits do not preserve the error budget of the fixed parameters, fast MCMC fitters treat the red noise by detrending, and sometimes the convergence cannot be reached at low SNR. Full fits with a Gaussian process and nested sampling deliver full parameter posteriors, evidence, and noise hyperparameter posteriors that we do not use, in a regime where the sampled shape parameters are not resolved, at a cost of a week per target. A full MCMC would explore broad, degenerate, multi-dimensional posteriors that are ultimately marginalized anyway to only get a planetary radius posterior that we need for population inference. Therefore, we simplify the problem rather than the sampler: we develop a method for low-SNR signals that exploits the parameter degeneracies and reduces the dimensionality of the parameter space, while preserving the error budget from all the latent parameters through pre-computed priors. The red noise is taken into account by whitening the light curves before folding them, using a whitening filter estimated from the data \citep[][]{ivashtenko_2025}; when fitting, we also apply a high-pass filter to both the folded data and the model to remove the power from the non-informative low frequencies that the finite data length does not allow to whiten properly.

\paragraph{This work in a broader context}
The same problem exists in other fields of astrophysics, where the parameter space is often multi-modal and more complex than for transiting planets. In gravitational-wave astronomy, re-parameterizing the 15-dimensional model space and semi-analytical marginalization resulted in orders-of-magnitude speed-up of the inference \citep[][]{islam_2022_factorized_pe, roulet_2022_multimodality, mushkin_2025_dot_pe}. In gravitational microlensing, the thousands of binary events expected from Roman motivated replacing per-event sampling by amortized inference \citep[][]{zhang_2021_microlensing_npe}, and even with fast fitters \citep[e.g.][]{eastman_2025_exozippy, pymc} the inference would benefit from an efficient parametrization \citep[][]{eastman_2026_sagan_talk}. The need will grow: PLATO \citep[][]{margin_2018_plato} is expected to yield $\sim5000$ exoplanets and Roman $\sim10^5$ transiting planets \citep[][]{wilson_2023}, and the number of candidates requiring parameter estimation will be even larger.
%, as the ratios of threshold crossing events to KOIs in Kepler \citep[][]{li_2019_dv_fit, thompson_2018} and to TOIs in TESS \citep[][]{guerrero_2021_toi} show.
% SRC: Guerrero 2021 Sect. 1: 55,281 TCEs from 27,822 two-minute targets, 2,241 TOIs.

% SRC: Eastman 2019 Sect. 23.3 (R_z < 1.01, T_z > 1000); Gazak 2012 (R < 1.1); Carter & Winn 2009 Sect. 4.2 (variance of normalized error on 10^4 simulations); Veitch 2015 and Romero-Shaw 2020 (P-P plots on injections); Cook 2006, Talts 2018 (SBC); Vasist 2023, Zhang 2021 (coverage of neural posteriors).
Finally, the validation of transit fits routinely consists of convergence diagnostics such as the Gelman-Rubin statistic \citep[][]{gazak_2012_tap, eastman_2019_exofast_2}, which test whether the sampler has reproduced its target distribution, not whether that distribution is calibrated with respect to the true parameters. Coverage tests on simulated transits have been done for specific questions \citep[][]{carter_2009} but are not routine, whereas in gravitational-wave parameter estimation, the percentile-percentile test on injections is standard \citep[][]{veitch_2015_lalinference, romero_shaw_2020_bilby}, as is simulation-based calibration in statistics and in simulation-based inference \citep[][]{cook_2006, talts_2018, zhang_2021_microlensing_npe, vasist_2023}. We adopt this practice and calibrate our posteriors on simulations (Section~\ref{sec:performance_simulation}).

\subsection{Structure of this paper}
In this work, we describe the method developed to derive the planetary radii posteriors, summarized in Section \ref{sec:summary_method}. In Section \ref{sec:mathematical_formulation}, we present the mathematical formulation of the method. Implementation details of each stage of the algorithm are provided in Section \ref{sec:method_details}. In Section \ref{sec:performance_simulation}, we characterize the performance of the method using simulated data. Then, in Section \ref{sec:comparison}, we compare the radii measured for faint Kepler planetary candidates with the values reported by Kepler. Finally, Section \ref{sec:discussion} discusses the limitations of this approach and the directions for future work.

\section{Methods}
\label{sec:methods}

\subsection{Summary of the method}
\label{sec:summary_method}

\begin{figure*}[h!t]
    \centering
    \includegraphics[width=0.9\textwidth]{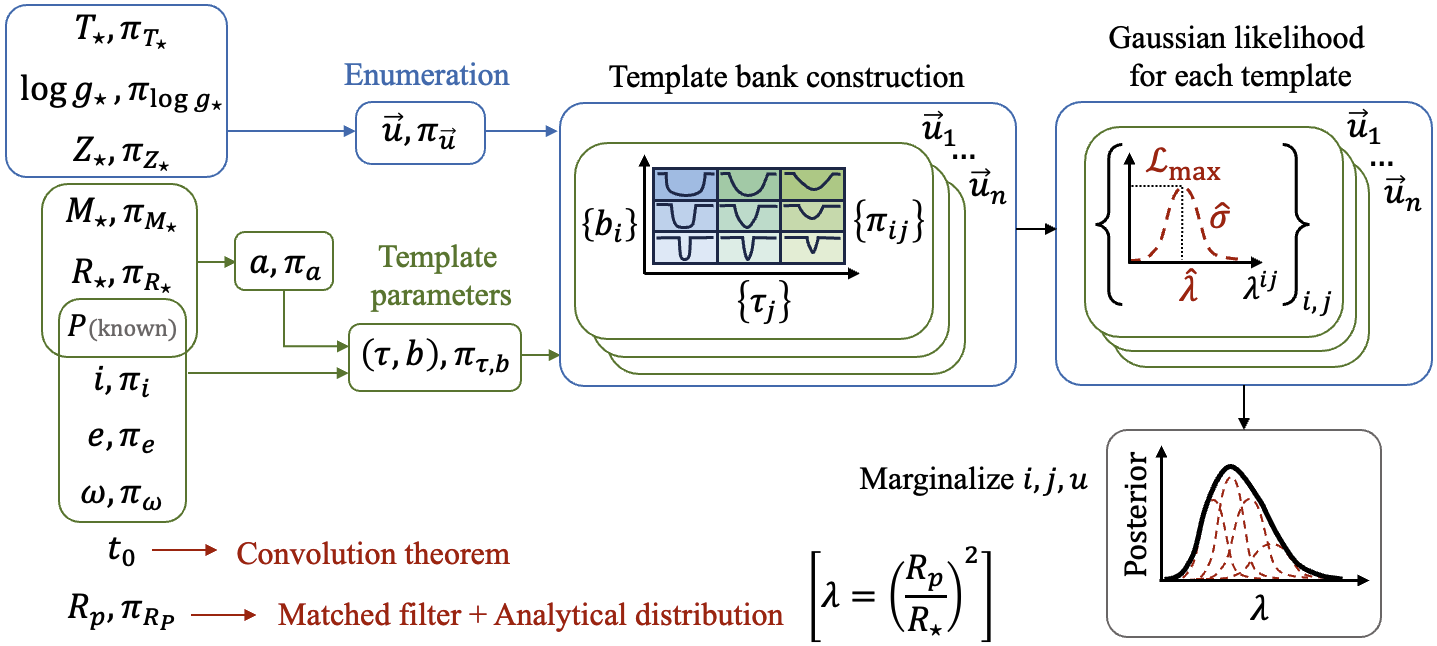}
    \caption{Schematic representation of the planetary radius posterior estimation algorithm. The left column shows the full set of system parameters taken with their priors. The arrows show how each parameter is treated in the algorithm, as described in the text.}
    \label{fig:full_method_scheme}
\end{figure*}

The method exploits symmetries of the transit model and properties of the likelihood to compute the posterior distribution of the planet-to-star radius ratio while avoiding sampling the full multi-dimensional parameter space. 

The full parameter space consists of 11 parameters with prior distributions: the orbital period $P$, epoch $t_0$, eccentricity $e$, periastron argument $\omega$, and inclination $i$; planetary radius $R_p$; and the stellar mass $M_{\star}$, radius $R_{\star}$, effective temperature $T_{\star}$, surface gravity $\log g_{\star}$, and metallicity $Z_{\star}$. Our goal is to compute the posterior distribution $p(\lambda|\mathbf{d})$ of $\lambda=(R_p/R_{\star})^2$ given the folded binned whitened flux $\mathbf{d}$. For efficiency, we want to call the transit model generator and evaluate the likelihood as few times as possible.

We treat each parameter according to how it enters the likelihood:
\begin{itemize}[noitemsep, topsep=0pt, left=1pt, label=-]
    \item $P$ is assumed known because it is measured very precisely from the periodicity search.
    \item $T_{\star}$, $\log g_{\star}$, $Z_{\star}$ and their priors are only used to produce the limb darkening coefficients $\mathbf{u}=(u_1,u_2)$ and their prior $\pi(\mathbf{u})$.
    \item $M_{\star}$ and $R_{\star}$ with their priors are only used to produce the prior distribution of semi-major axis $a$, $\pi(a)$.
    \item $\lambda$ is treated semi-analytically using the approximate linearity of the transit model.
    \item $t_0$ is treated using the convolution theorem.
    \item $e, \omega, i, a$ and their priors are used to construct a template bank parametrized by transit duration $\tau$ and impact parameter $b$. This parametrization is natural because, for the observer, $e, \omega, i, a$ determine how close to the limb ($b$) and how fast ($\tau$) the transit is. Our template bank covers the same transit shapes as the original parameter space and preserves the original probability mass using the prior $\pi(\tau, b)$. 
\end{itemize}
As a result, the transit model is computed only for the templates, and the enumeration runs over the 4-dimensional space $(\tau, b, \mathbf{u})$ instead of the 11 original parameters, while the template priors preserve the probability mass and hence the marginalized $\lambda$ posterior. A scheme summarizing how the parameters are treated is shown in Figure~\ref{fig:full_method_scheme}.

% -----------------------------------------------------------------
\subsection{Mathematical formulation of the method}
\label{sec:mathematical_formulation}
We first summarize the method in one equation and then expand it.

We denote the vector of the nuisance parameters as $\boldsymbol{\theta}$, with prior $\pi(\boldsymbol{\theta})$. It contains all parameters except for $\lambda=(R_p/R_{\star})^2$, and we also assume 
$\pi(\lambda, \boldsymbol{\theta})=\pi(\lambda)\pi(\boldsymbol{\theta})$. The symmetries and properties of the likelihood allow us to reduce the problem to an effective parameter vector $(\tau, b, \mathbf{u})$ that we denote $\boldsymbol{\eta}$. Our method computes the $\lambda$ posterior as
\begin{align}
    p(\lambda|\mathbf{d})
    &=  
    \frac{1}
    {\mathcal{L}(\mathbf{d})} 
    \! \int \! \!
    d\boldsymbol{\theta}
    \pi(\lambda, \boldsymbol{\theta})
    \mathcal{L}(\mathbf{d}|\lambda, \boldsymbol{\theta}) 
    \label{eq:summary_equation_initial}
    \\&\approx
    \frac{\pi(\lambda)}
    {\mathcal{L}(\mathbf{d})}
    \sum_k \pi(\boldsymbol{\eta}_k)
    \mathcal{L}_{\hat{\lambda}}(\mathbf{d}|\boldsymbol{\eta}_k)
    \exp{\left(\!\!-\frac{(\lambda\!-\!\hat{\lambda}_k)^2}{2\sigma^2_{\lambda,k}}\!\right)}
    ,
    \label{eq:summary_equation_final}
\end{align}
where 
\begin{align}
\begin{split}
\pi(\boldsymbol{\eta}_k)=
\int_{\boldsymbol{\theta}\in \boldsymbol{\Theta}[\boldsymbol{\eta}_k]}
\!\!\!\!\!\!\!d \boldsymbol{\theta}
\pi(\boldsymbol{\theta}),
\label{eq:template_prior}
\end{split}
\end{align}
and $\mathcal{L}_{\hat{\lambda}}(\mathbf{d}|\boldsymbol{\eta}_k)$ is the likelihood at the maximum-likelihood estimate $\hat{\lambda}$.
The transit epoch is marginalized by inflating $\sigma_{\lambda,k}^2$ with the variance of the epoch-dependent estimates $\hat{\lambda}_k(t_0)$ around the best-fitting epoch (Appendix~\ref{app:t0_marginalization}).

Below, we elaborate on this approximation mathematically; Section~\ref{sec:method_details} describes its implementation.

\subsubsection{Degenerate parameters dimensionality reduction}
\label{sec:math_template_bank}

In Equation \ref{eq:summary_equation_final}, we use the fact that the transit model has symmetries, meaning that different $\boldsymbol{\theta}$ may give nearly identical models. This allows us to construct a lower-dimensional, discretized effective parameter vector $\boldsymbol{\eta}$ and compress the probability mass of $\boldsymbol{\theta}$ into the prior $\pi(\boldsymbol{\eta})$:

\begin{align}
    \int d\boldsymbol{\theta}
    \mathcal{L}(\mathbf{d}|\lambda, \boldsymbol{\theta}) 
    \pi(\lambda, \boldsymbol{\theta})
    \approx
    \sum_k \pi(\boldsymbol{\eta}_k)
    \mathcal{L}(\mathbf{d}|\lambda, \boldsymbol{\eta}_k).
    \label{eq:likelihood_template_bank}
\end{align}
Equation~\ref{eq:template_prior} defines the template prior, where $\boldsymbol{\Theta}[\boldsymbol{\eta}_k]$ is the set of all $\boldsymbol{\theta}$ for which $\boldsymbol{\eta}_k$ gives the closest matching transit model. 

Section~\ref{sec:template_bank_construction} elaborates on the choice of the template parameters, explains how the templates are generated, and the precision of the approach.

% --------------------------------------------------
\subsubsection{Planet-to-star radius ratio semi-analyticity}
\label{sec:math_planetary_radius}

Consider the transit model $\mathbf{s}(\boldsymbol{\theta}, \lambda)$ that is a vector indexed by time. The model scales linearly with $\lambda$ since the planetary radius squared defines the transit depth. In addition, $\lambda$ also impacts the ingress and the egress of the transit. Using the small-planet approximation, we assume that the ingress-egress effect is much weaker and more slowly varying than the linear scaling, so that the model can be written as 
\begin{equation}
\mathbf{s}(\boldsymbol{\theta}, \lambda)
\approx\frac{\lambda}{\lambda_c}\mathbf{s}(\boldsymbol{\theta}, \lambda_c)
=\lambda\mathbf{h}(\boldsymbol{\theta}, \lambda_c),
\label{eq:lambda_adiabatic}
\end{equation}
where $\lambda_c$ is a coarse value of $\lambda$ that sets the ingress-egress effect, while the precise $\lambda$ fine-tunes the transit depth. We refer to all the changes in the transit model apart from depth scaling as transit shape. Thus, $\mathbf{h}$ fixes the transit shape, and $\lambda$ scales the model to the correct depth.

We can then write the statistical model of the whitened folded binned data vector as
\begin{align}
    \mathbf{d}\sim \lambda \mathbf{h}
    %(\boldsymbol{\theta}, \lambda_c)
    +\mathcal{N}\left(0,\boldsymbol{\sigma}\right),
    \label{eq:data_model}
\end{align}
where $\boldsymbol{\sigma}$ is a time-dependent standard deviation of the folded whitened data, and the mean was subtracted. The original whitened light curve has unit variance, but the folded and binned light curve may have different variance in every bin due to unequal number of data points contributing to this bin.

The likelihood for this statistical model can be written as
\begin{align}
    \mathcal{L}(\mathbf{d}|\lambda, \boldsymbol{\theta})
    &\propto
    \exp\!\left[-\frac{1}{2}
    \left\|\frac{ \mathbf{d} - \lambda\mathbf{h} }
    {\boldsymbol{\sigma}}\right\|^2
    \right]
    \label{eq:original_likelihood}
    \\&\propto
    \mathcal{L}_{\hat{\lambda}}
    \exp\!\left[-\frac{(\lambda-\hat\lambda)^2}{2\sigma_{\lambda}^2}\right]
    ,
    \label{eq:factorized_likelihood_lambda_hat}
\end{align}
where
\begin{align}
    \hat{\lambda}
    =\frac{\left<\mathbf{d}, \mathbf{h}\right>}
    {\left<\mathbf{h}, \mathbf{h}\right>},
    \label{eq:amplitude_estimator}
\end{align}
is the maximum-likelihood estimate for $\lambda$ \citep[][]{kay_1993}. The inner product is 
\begin{align}
    \left<\mathbf{d}, \mathbf{h}\right>
    =\sum_i v_i d_i h_i.
    \label{eq:inner_product}
\end{align}
with weights $v_i = 1/\sigma_i^2$. This estimate has a Gaussian distribution with variance
\begin{align}
    \sigma_{\lambda}^2 =  
    \frac{1}
    {\left<\mathbf{h}, \mathbf{h}\right>}.
    \label{eq:sigma_amplitude_estimator}
\end{align}
$\mathcal{L}_{\hat{\lambda}}$ is the maximal likelihood corresponding to this estimate,
\begin{equation}
    \mathcal{L}_{\hat{\lambda}}
    \propto
    \exp\!\left[-\frac{1}{2}
    \left\|\frac{ \mathbf{d} - \hat{\lambda}\mathbf{h} }
    {\boldsymbol{\sigma}}\right\|^2
    \right].
    \label{eq:max_likelihood}
\end{equation}
Thus, Equation~\ref{eq:factorized_likelihood_lambda_hat} shows that the likelihood as a function of $\lambda$ can be computed analytically. It is a Gaussian with the mean at the maximum-likelihood $\hat{\lambda}$ scaled by the corresponding likelihood $\mathcal{L}_{\hat{\lambda}}$.

\paragraph{Marginalization}

We can thus compute the full marginalized posterior for $\lambda$ analytically as a Gaussian mixture, summing the Gaussian likelihoods (Equation~\ref{eq:factorized_likelihood_lambda_hat}) corresponding to all the templates with $\boldsymbol{\eta}_k$ (Equation~\ref{eq:likelihood_template_bank}). In this way, we get Equation \ref{eq:summary_equation_final}.

The marginalization over the transit epoch $t_0$ is done when computing $\mathcal{L}_{\hat{\lambda}}$, as described in Section~\ref{sec:planetary_radius}.
% Since the $\lambda$ estimator (Equation~\ref{eq:amplitude_estimator}) is an inner product, computing it in the Fourier domain gives the result for any discrete time shift, with no need to iterate over $t_0$ and generate new models. This marginalization is not significant for matched-filter SNR of the folded signal $>7$ and is elaborated in Appendix~\ref{app:t0_marginalization}.

According to Equation~\ref{eq:lambda_adiabatic}, we still need to choose and approximate $\lambda_c$ to generate a transit template for any $\boldsymbol{\eta}_k$. We do this by first estimating the relative transit depth $\delta$ from the data and then inverting the equation for the transit depth from \citet[][]{csizmadia_2013},
\begin{equation}
    \lambda_c(b, \mathbf{u})
    = \delta
    \frac{1-u_1/3 - u_2/6}
    {1 - u_1(1-\mu) - u_2(1-\mu)^2},
\end{equation}
where $\mu = \sqrt{1-b^2}$. Since $\mathbf{u}$ and $b$ are among the parameters $\boldsymbol{\eta}$, we can generate this $\lambda_c$ guess for each template.

% \begin{equation}
%     p(\lambda|d)\propto
%     \mathbf{1}_{[\lambda_{\min},\lambda_{\max}]}(\lambda)
%     \sum_j \pi_j \mathcal{L}_{\hat{\lambda},j}
%     \exp\!\left[-\frac{(\lambda-\hat\lambda_j)^2}{2\sigma_{\lambda,j}^2}\right],
%     \label{eq:mixture}
% \end{equation}

% Section~\ref{sec:method_details} elaborates how this formalism was implemented in our algorithm, how $\boldsymbol{\theta}$ parameters were 

\subsection{Details of the algorithm}
\label{sec:method_details}

This section elaborates on how the formalism of Section~\ref{sec:mathematical_formulation} was implemented in our algorithm. 
% Subsection~\ref{sec:stellar_parameters} explains the provenance of stellar parameters and their priors. Then, Subsection~\ref{sec:limb_darkening_coefficients} explains how the limb darkening coefficients were sampled. Subsection~\ref{sec:template_bank_construction} explains the choice of the template parameters and how they were generated. Subsection~\ref{sec:planetary_radius} provides more details on the planetary radius treatment. Subsection~\ref{sec:target_preparation} describes how the whitened folded binned data was prepared. Finally, Subsection~\ref{sec:inference process} provides details of how the inference process was performed.

This work is designed for the cases of low-SNR (matched-filtering SNR of the folded whitened signal $\lesssim 20$), periods between about 30 and 500 days, and with no TTVs, corresponding to our search pipeline \citep[][]{ivashtenko_2025}, although it can be used beyond these limits. We condition on the signal being a small planet, i.e. $b<1$, and no eclipsing-binary contamination. A detailed list of method assumptions is provided in Appendix~\ref{app:assumptions}.

\subsubsection{Stellar parameters}
\label{sec:stellar_parameters}
We use stellar radius, effective temperature, surface gravity, and metallicity from the Gaia DR3 GSP-Phot, and stellar mass from Gaia DR3 FLAME \citep[][]{gaia_2022_dr3}. For targets for which Gaia data are not available, we use the stellar properties from \citet[][]{berger_2020}. 
% We prioritize Gaia because of uniformity

We need distributions of the stellar parameters, whereas the catalogs report only a value with asymmetric uncertainties.
The catalog uncertainties are moreover known to be underestimated, as they do not include systematic errors \citep[][]{andrae_2023_gaia_bp_rp_analysis}. To quantify the missing uncertainty, we compare determinations of the same parameters for $\sim8\times10^4$ overlapping FGK Kepler targets: the GSP-Phot stellar radii against the FLAME and \citet[][]{berger_2020} values, and the Gaia masses, effective temperatures, metallicities, and surface gravities against those of \citet{berger_2020}. The differences are broader than the reported uncertainties allow, with a median absolute difference of $\sim$1.1--1.5 times the combined reported error, where 0.6745 is expected for consistent Gaussian errors.

For each parameter, we therefore derive a systematic error floor $\kappa$: the additional uncertainty added in quadrature to the reported errors so that the median of $|\Delta|/\sqrt{\sigma_\Delta^2+\kappa^2}$ equals 0.6745, where $\Delta$ is the difference between the two catalog values and $\sigma_\Delta$ their combined reported error. 
% Using the median makes the estimate robust to the non-Gaussian tails of the difference distributions.

% We solve for $\kappa$ in the natural space of each parameter. The mass,
% radius, and effective temperature systematics are multiplicative, so we
For mass, radius, and effective temperature we use logarithmic differences of the catalog values $v_1$, $v_2$ in two catalogs, $\Delta = \ln(v_2/v_1)$, and $\sigma_\Delta^2 = (\sigma_1/v_1)^2 + (\sigma_2/v_2)^2$.
% , in which a relative error is an additive floor; the resulting $\kappa$ is the relative error. 
Metallicity and $\log g_\star$ are already logarithmic quantities, so we use absolute differences and obtain $\kappa$ in dex.
% For the multiplicative parameters, solving in linear space returns the same values, confirming the consistency of this description.
As a result, we add a 6\% systematic error for the stellar radius, 4\% for the mass, and 4\% for the effective temperature, and absolute floors of 0.27~dex for the metallicity and 0.06~dex for $\log g_\star$. 
We add them in quadrature to the reported statistical uncertainties of the catalog used for each target. 
% This conservatively attributes the entire cross-catalog discrepancy to the catalog in use.

To get priors of each stellar parameter for our inference, we fit a skew-normal distribution to the catalog median and inflated 16th and 84th percentiles, and then draw a stratified sample from these distributions. 

\paragraph{Period and semi-major axis}
We take the orbital period as exactly known from the periodicity search (Section~\ref{sec:intro}).

The transit model does not depend explicitly on stellar mass $M_{\star}$ and radius $R_{\star}$; it depends only on the scaled semi-major axis $a/R_{\star}$. Therefore, we reduce $M_{\star}$ and $R_{\star}$ priors to the distribution of $a/R_{\star}$. We draw $M_{\star}$ and $R_{\star}$ samples from their priors, convert them to $a/R_{\star}$ using Kepler's third law and construct a histogram of this sample to get the prior for $a/R_{\star}$.
It will be used to construct the template bank prior (Section~\ref{sec:template_bank_construction}).

\subsubsection{Limb darkening coefficients}
\label{sec:limb_darkening_coefficients}

Stellar metallicity, surface gravity, and effective temperature enter the transit model only through the limb darkening coefficients (LDC). We use the quadratic limb darkening law with coefficients $\mathbf{u}=(u_1,u_2)$ tabulated for the Kepler bandpass by \citet{sing_2010_limb_darkening}. We sample the stellar parameters from their priors (Section~\ref{sec:stellar_parameters}) and get LDC for each sample from the interpolated table, propagating the stellar parameter uncertainty. 

This propagation alone is known to underestimate the LDC uncertainty \citep[][]{csizmadia_2013, howarth_2011, patel_2022, maxted_2018}, 
% The differences between theoretical tables, and between tabulated and empirically measured coefficients, can reach $\sim0.2$ \citep[][]{csizmadia_2013, howarth_2011, patel_2022}. Figure~\ref{fig:ldc_sampling} demonstrates the LCD scatter due to the stellar parameters errors, compared to the full scatter
% in \citet[][]{csizmadia_2013} for their stellar parameter errors (ours are larger though) it contributes only $\lesssim0.03$ to the coefficients
therefore we add scatter around every $\mathbf{u}$ for each stellar-parameter sample. 
The measured $u_1$ and $u_2$ are known to be strongly anti-correlated \citep[][]{howarth_2011}, with a correlation coefficient between $-0.75$ and $-0.99$ depending mostly on the impact parameter \citep[][]{pal_2008_ldc_correlation}. We describe the correlated scatter in the basis aligned with the principal axes of the correlation ellipse,
\begin{align}
\begin{split}
    \gamma_1 &= u_1\cos\phi - u_2\sin\phi,
    \\ \gamma_2 &= u_1\sin\phi + u_2\cos\phi,
\end{split}
\end{align}
with $\phi=40^\circ$. This places the poorly constrained direction $\gamma_1$ at $-40^\circ$ from the $u_1$ axis, within $\sim 10^\circ$ of the decorrelation direction of \citet{pal_2008_ldc_correlation}, who recommends $\phi\sim40^\circ$ in a convention rotated oppositely to ours.
%for the case when the impact parameter is not known in advance; 
In this basis, we draw uncorrelated Gaussian scatter with the larger standard deviation along $\gamma_1$, $\sigma_{\gamma_1} = 0.4\max(\gamma_1,\gamma_2)$, which evaluates to
$\approx0.2$ for typical FGK targets, matching the observed offsets between tabulated and empirically measured coefficients \citep{patel_2022, csizmadia_2013, maxted_2018}. The relative scaling reflects the degradation of the theoretical predictions toward cooler stars, whose coefficients are larger \citep{csizmadia_2013, patel_2022}. The standard deviation along $\gamma_2$ follows from the correlation ellipse axis ratio $\sqrt{(1+|\rho|)/(1-|\rho|)}$, where we take $|\rho|=0.7$, conservatively widening the predicted range of \citet{pal_2008_ldc_correlation}. 
%  widens the well-constrained direction to $\sigma_{\gamma_2}\approx0.08$, above the $\sim0.01$--$0.05$ accuracy demonstrated for the tightly measurable combination of the coefficients \citep{maxted_2018}
% conservatively widens the well-constrained direction to $\sigma_{\gamma_2}\approx0.08$, above the $\sim0.01$--$0.05$ accuracy demonstrated for the tightly measurable combination of the coefficients \citep{maxted_2018}. 
We note that for stars cooler than $\sim5000$~K the theory-observation offsets can exceed these values \citep{patel_2022}, but such stars are rare in our sample. We also note that the LDC contribution to the radius error budget is subdominant (Appendix~\ref{app:error_budget_parameters}), so these choices have little effect on the final $R_p/R_{\star}$ posterior.

\begin{figure}[h]
    \centering
    \includegraphics[width=0.47\textwidth]{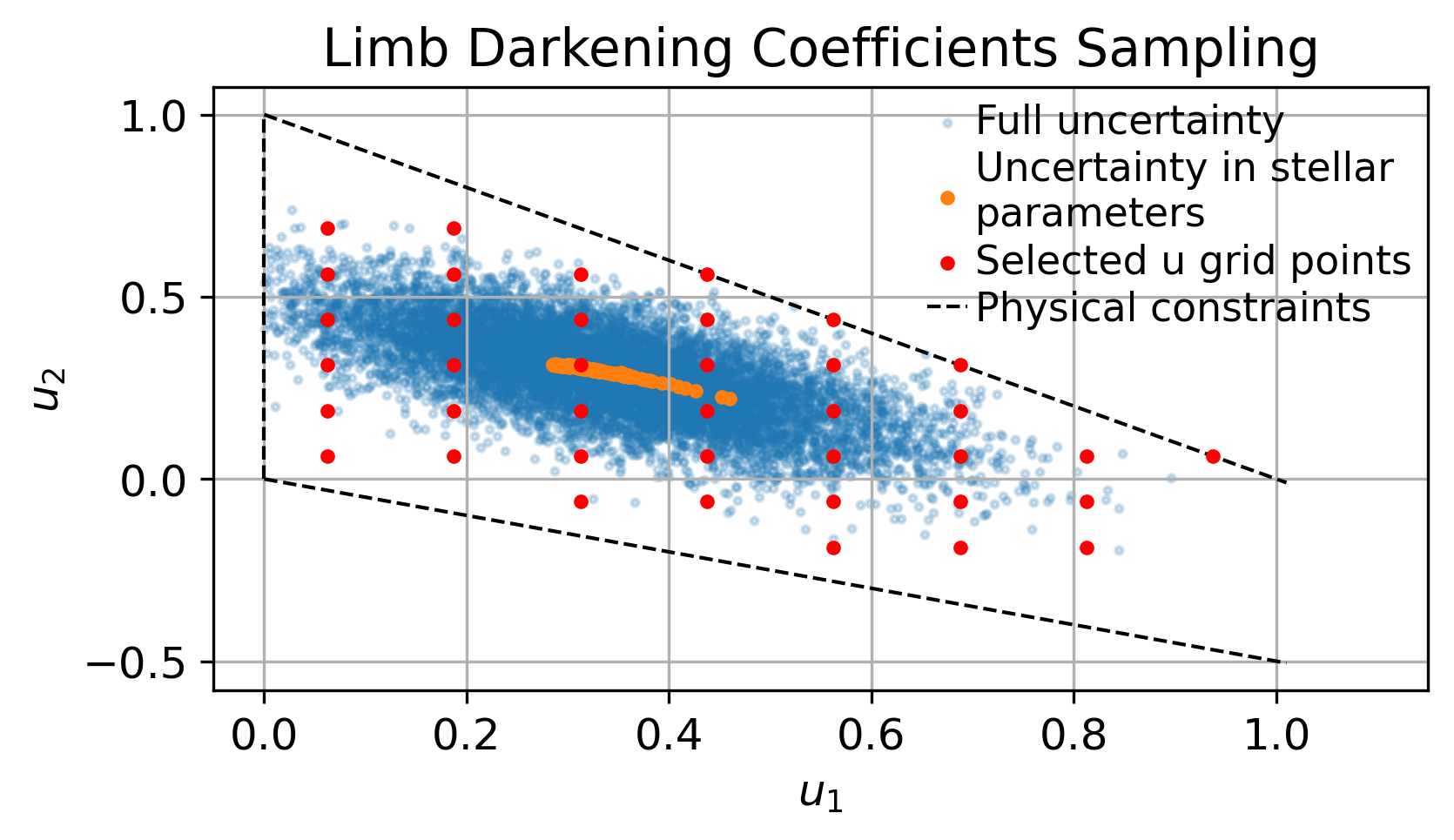}
    \caption{Illustration of the limb darkening coefficients $u_1$, $u_2$ sampling. The orange points denote the uncertainty in the coefficients coming from the stellar parameter uncertainty. The remaining variance is caused by the empirical $u_1$, $u_2$ uncertainty for fixed stellar parameters. The dashed lines show the physical constraints on the coefficients (Equation~\ref{eq:ldc_constraints}) \citep[][]{kipping_2013_ldc_sampling}. The red dots are the grid points that will be used for enumeration during the fit.}
    \label{fig:ldc_sampling}
\end{figure}

Finally, we rotate the coefficients back to the $u_1, u_2$ plane and apply the physical LDC constraints \citep[][]{kipping_2013_ldc_sampling},
\begin{align}
\begin{split}
    & u_1+u_2 \leq 1,
    \\ & u_1 + 2u_2 \geq 0,
    \\ & u_1 \geq 0.
    \label{eq:ldc_constraints}
\end{split}
\end{align}
The obtained distribution is illustrated in Figure~\ref{fig:ldc_sampling}, highlighting the uncertainty due to the stellar parameters distribution and the additional empirical error. We select a grid of points that will be used for enumeration, as shown with red dots in Figure~\ref{fig:ldc_sampling}. For every grid cell, we compute the prior by summing the weight of the samples in this cell.
During the fit, we will enumerate over all the grid cells and add the corresponding priors to the likelihoods, as shown in Figure~\ref{fig:full_method_scheme}.
               
\subsubsection{Template bank construction}
\label{sec:template_bank_construction}

For the dimensionality-reduced Equation~\ref{eq:likelihood_template_bank} to yield a speedup, evaluating the likelihood of a template must not be significantly more expensive than that of the original parameters, and $\pi(\boldsymbol{\eta})$ must be cheap to compute from $\pi(\boldsymbol{\theta})$. We therefore choose $\boldsymbol{\eta}$ such that the transit model is generated directly by a standard integrator \citep[e.g.][]{kreidberg_2015_batman}, and $\boldsymbol{\eta}$ is an analytic function of $\boldsymbol{\theta}$.

\paragraph{Orbital element parametrization}
The transit model \citep[][]{kreidberg_2015_batman} accepts as input limb-darkening coefficients $\mathbf{u}$, orbital period $P$ (fixed), semi-major axis $a$ in units of stellar radii $R_\star$, eccentricity $e$, periastron argument $\omega$, inclination $i$, and planetary radius as $R_p/R_{\star}$. With given $\mathbf{u}$, we seek to re-parametrize the orbital parameters $a$, $e$, $\omega$, $i$.

We find that a two-parameter family of circular-orbit models (the template bank) contains a sufficiently close match to the transit model of almost any orbit. This bank can be parametrized by approximate analytical values of transit duration $\tau$ and impact parameter $b$ for every full orbital parameter set:
\begin{align}
    b & = \frac{a (1 - e^2) \cos i}{1 + e \sin \omega},
    \label{eq:b}
    \\ v &= \frac{2 \pi a
    \sqrt{ 
    1 + 2e \sin \omega
    + e^2 - e^2 \sin^2 i \cos^2 \omega
    }}
    {P \sqrt{1-e^2}} ,
    \label{eq:v}
    \\ \tau &= \frac{2 \sqrt{1 - b^2}}{v},
    \label{eq:tau}
\end{align}
where $v$ is the planet velocity in transit. We choose this parametrization because $\tau$ and $b$ can be easily computed analytically, and can be used to directly produce a transit model, as shown below. A disadvantage of this choice is that $\tau$ and $b$ are strongly correlated. An alternative way could be to use principal component analysis to find a non-degenerate parametrization numerically and generate non-physical effective models \citep[as done in][]{roulet_2019_template_bank}.

We note that we derive the expressions~\ref{eq:b}-\ref{eq:tau}  based on the assumptions listed in Appendix~\ref{app:assumptions}. Different simplifying assumptions lead to slightly different expressions in the literature \citep[][]{van_eylen_2019, seager_2003_transit_parameters, carter_2008}.

\paragraph{Template prior computation}
We construct a grid of $\log \tau$ and $b$ that will be used to generate the transit models for the fit. The full orbital parameter distribution is used to compute prior probabilities for this grid. We iterate over semi-major axis $a$ with its prior distribution obtained in Section~\ref{sec:stellar_parameters}. 
For each $a$, we draw $e$ and $\omega$ from a prior uniform in $(\sqrt{e} \cos\omega, \sqrt{e} \sin\omega)$ \citep[e.g.][]{van_eylen_2019} and $\cos i$ uniform on $[0, 1]$, and retain only the configurations that transit. The retained sample is thus distributed according to the orbital prior conditioned on transit. For every ($a$, $e$, $\omega$, $i$) set, we calculate analytically $(\log\tau, b)$ using Equations~\ref{eq:b},\ref{eq:tau} and add the prior weight to the corresponding cell of the $(\log\tau, b)$ table. 

\paragraph{Template generation}

To generate a transit model, we need to provide parameters that the integrator accepts. We convert $\tau$ and $b$ to effective values for semi-major axis and inclination $a^e$, $i^e$ for a circular orbit:

\begin{align}
    a^e &= \frac{\sqrt{1 - b^2} P}{\pi \tau},
    \label{eq:a_e}
    \\ \cos i^e &= b/a^e.
    \label{eq:iota_e}
\end{align}
These values are not physical; they only serve to generate transit models of the required shapes. 

\paragraph{Parametrization efficiency}
The template bank covers the space of transit shapes only approximately, both because it is discretized and because of the parametrization itself. The simplifying assumptions (Appendix~\ref{app:assumptions}) exclude orbital configurations with extreme eccentricities and periastron angles; Equations~\ref{eq:b},\ref{eq:tau} are only approximate; and the effective templates obtained with Equations \ref{eq:a_e}, \ref{eq:iota_e} may not match the desired shapes exactly. To verify the coverage of the template bank, we generate transit models with full orbital parameter set, calculate their two-dimensional parametrization (Equations~\ref{eq:b},\ref{eq:tau}), match them to the corresponding template from the bank, and compute the inferred planetary radius from the best-fit amplitude. The distributions of the obtained errors are shown in Appendix~\ref{app:tb_errors}. From the prior-weighted transiting planet parameter set, and with the grid density used in the fit, in 95\% of the cases the error is below 3\%. The large error cases correspond to eccentricities $e>0.9$.

The transit depth is not set by $R_p/R_\star$ alone; it also depends on the impact parameter and the limb darkening  \citep[e.g.][]{csizmadia_2013}. Since $b$ is one of the template parameters, this dependence is largely captured. Neglecting this would lead to systematically overestimating radii for small $b$ and underestimating for large $b$. In addition, we use the $(b,\mathbf{u})$-informed initial guess on the planetary radius to generate the templates (Section~\ref{sec:mathematical_formulation}). This captures the ingress and egress changes due to the planetary radius.

\subsubsection{Planetary radius}
\label{sec:planetary_radius}

\paragraph{Transit epoch}
The data are folded with the known period, and the transit epoch $t_0$ is the position of the transit within the folded window. As mentioned in Section~\ref{sec:math_planetary_radius}, we compute the estimator (Equation~\ref{eq:amplitude_estimator}) in the Fourier domain, which by the convolution theorem yields $\hat{\lambda}(t_0)$ for every integer bin shift. 
We then marginalize over $t_0$ in a window around the maximum-likelihood epoch, with a uniform prior on $t_0$, by replacing the Gaussian mixture over epochs with a single Gaussian of the same variance centered at the maximum-likelihood amplitude (Appendix~\ref{app:t0_marginalization}). 
Epochs at which the estimated amplitude is negative (anti-transits) are excluded from the window, consistently with the prior $\lambda\geq0$. 

In practice, above the detection threshold of our search \citep[matched-filter SNR$>7$ for the folded whitened signal;][]{thompson_2018, ivashtenko_2025}  the weights are dominated by the maximum-likelihood epoch, and marginalization is nearly identical to maximization over $t_0$. 

A non-integer shift in $t_0$ (sub-bin shift) cannot be handled by the convolution theorem. It is negligible for transits much longer than a bin; for short transits it changes the binned transit shape, and in such cases we compute the model at several sub-bin shifts.

\paragraph{Mask-aware Fourier domain matched filter}
The matched-filtering estimator (Equation~\ref{eq:amplitude_estimator}) can be computed in the Fourier domain even if the whitened folded binned flux has time-dependent variance or a mask filtering out several bins. This arises because unequal numbers of folded flux points contribute to each bin. In extreme cases, such as a small number of transits or a period close to an integer multiple of the cadence, some bins can remain completely empty and need to be masked.

We include a mask vector $\mathbf{m}$ with zeros on bad bins and one on good bins in the weight vector in the inner product Equation~\ref{eq:inner_product}, $\mathbf{v}=\mathbf{m}/\boldsymbol{\sigma}^2$, where $\boldsymbol{\sigma}^2$ is the per-bin variance, and vector division means per-element division. 

We then compute the numerator and the denominator of Equation~\ref{eq:amplitude_estimator} for any integer shift $(t)$ using the convolution theorem in the Fourier domain, 
\begin{align}
    &\left<\mathbf{d}, \mathbf{h}_t\right>(t) 
    =\mathcal{F}^{-1}
    \left[
    \mathcal{F}[\mathbf{d} \circ \mathbf{v}]
    \circ
    \mathcal{F}^{*}[\mathbf{h}]
    \right],
    \\&\left<\mathbf{h}_t, \mathbf{h}_t\right>(t) = 
    \mathcal{F}^{-1}
    \left[
    \mathcal{F}[\mathbf{v}]
    \circ
    \mathcal{F}^{*}[\mathbf{h} \circ \mathbf{h}]
    \right],
\end{align}
where $\mathcal{F}$ and $\mathcal{F}^{-1}$ denote the Fourier transform and the inverse Fourier transform, asterisk denotes complex conjugation, and $\circ$ denotes per-element multiplication. The $\left<\mathbf{h}, \mathbf{h}\right>(t)$ term is now a vector and not a scalar as it would be without $\mathbf{v}$, since it needs to be computed for every shift between the model $\mathbf{h}$ and the weight vector. Still, computing it has the same computational complexity as computing $\left<\mathbf{d}, \mathbf{h}\right>$, so the cost is not significantly increased.

For the cases where bin-wise variances are similar and there are no masked bins in the area where the whitened transit power is concentrated, we perform the fit without weights.

\paragraph{Prior distribution of $\lambda$}
Since the likelihood is Gaussian in $\lambda$ with a $\lambda$-independent width, its Jeffreys prior is the improper uniform prior (uniform prior whose integral diverges). We use a prior flat in $\lambda$ but bounded, in order to avoid diverging weights. 

For real targets, we set the upper bound of the prior in a way that it has no significant influence on the posterior shape. The lower bound at zero truncates the posterior mass at negative amplitudes, which matters only at very low SNR.

In our simulation (Section~\ref{sec:performance_simulation}), we use a narrow prior, since we also use it to sample the true values from it for calibration.

As an alternative to uniform priors, it is possible to use the currently known occurrence rates, e.g. \citet[][]{zhu_dong_2021}. However, the occurrence rates are not known precisely enough to change the inference significantly in most cases. In principle, since we plan to use the derived posteriors for re-evaluating occurrence rates, a more accurate approach would be to estimate the population radii distribution self-consistently. In this approach, the prior is iteratively adjusted and used to re-evaluate the resulting occurrence until the two converge.

\paragraph{Marginalization over $\lambda$}

As prescribed by Equation~\ref{eq:summary_equation_final}, for every template in the grid and every $\mathbf{u}$, we save the mean and variance of $\hat{\lambda}$ (Equations~\ref{eq:amplitude_estimator},~\ref{eq:sigma_amplitude_estimator}), and the likelihood value for the best-fit amplitude. To obtain the final $\lambda$ posterior, we marginalize over all other parameters by summing normalized Gaussian survival functions $S_G$ with the corresponding likelihood weights and the corresponding priors over all $\tau, b, \mathbf{u}$. 

For the improper flat prior in $\lambda$, the final survival function $S_0(\lambda)$ is
\begin{align}
\begin{split}
    S_0(\lambda) = \frac{\sum_k w_k 
    S_G\left(\lambda;\hat{\lambda}_k, \sigma_{\lambda,k}\right) }
    {\sum_k w_k},
    \label{eq:survival_function_no_prior}
\end{split}
\end{align}
where the weights are
\begin{align}
\begin{split}
    w_k = \pi_k\mathcal{L}_{\hat{\lambda}, k} \sigma_{\lambda,k}.
    \label{eq:survival_function_summation_weights}
\end{split}
\end{align}
The index $k$ runs over all the combinations of $\tau, b, \mathbf{u}$ that were used to produce templates, and $\pi_k$ is the prior mass for each of them. $\mathcal{L}_{\hat{\lambda}, k}$ is the best-$\lambda$ likelihood defined in Section~\ref{sec:mathematical_formulation}.

For our case of uniform bounded prior over $[\lambda_{\text{min}}, \lambda_{\text{max}}]$, the survival function is obtained from $S_0$ as
\begin{align}
\begin{split}
    S_\pi(\lambda) = \frac{S_0(\lambda) - S_0(\lambda_{\text{max}}) }
    {S_0(\lambda_{\text{min}}) - S_0(\lambda_{\text{max}})}.
    \label{eq:lambda_survival_function_with_prior}
\end{split}
\end{align}
This survival function can be computed for an arbitrary value of $\lambda$ to obtain quantiles, or differentiated on a grid to obtain the marginalized posterior. 

This is only a representation of the result: the algorithm itself never grids or samples $\lambda$.

\subsubsection{Target preparation}
\label{sec:target_preparation}

We perform the parameter estimation on folded whitened binned light curves, after masking invalid cadences. An example of such a light curve for one of the targets, together with the corresponding best-fit template, is shown in Appendix~\ref{app:inference_process_illustration}.

We use the long-cadence Simple Aperture Photometry after Pre-search Data Conditioning (PDCSAP) flux \citep[][]{pdc_kdph} from Kepler DR 25 \citep[][]{kepler_dr_notes_25} available on the MAST portal: 
% \url{doi:10.17909/T9488N}.
\dataset[doi:10.17909/T9488N]{https://doi.org/10.17909/T9488N}.
Our pre-processing performs several first steps of the search pipeline described in \citet[][]{ivashtenko_2025}, including mean subtraction, outlier rejection, bad cadence masking, normalization, low-frequency trend subtraction, and whitening. Folding is performed after whitening. The whitening filters of the quarters of the data are averaged to create an effective filter that is used to whiten the templates used for fitting. It is important to whiten both the data and the model before matching them \citep[][]{li_2019_dv_fit, ivashtenko_2025}, otherwise the result can be biased. We note that using folded data instead of the full ephemeris carries information loss due to differences between the whitening filters of different quarters. In Appendix~\ref{app:folded_data_loss}, we discuss this effect and show that it is not significant for our target sample (\ref{sec:comparison}).

For binning, we use a bin width of half the Kepler cadence to better resolve the transit shape. We upsample the average whitening filter accordingly. We account for heteroscedastic variances in the individual bins resulting from uneven number of measurements in each bin.

In order to avoid fit bias due to absent power in these frequencies, we apply a high-pass filter on both the data and the templates before the fit. Specifically, we use tapering with Tukey window applied both in Fourier and in time domain to make the filter have finite response length. We attenuate frequencies below $\approx0.38$ $\text{day}^{-1}$, i.e. timescales longer than 128 Kepler long cadences ($\approx2.6$ days).

The preprocessing (whitening, high-pass filtering, binning) can itself introduce errors, on which posteriors are usually implicitly conditioned. We estimated its potential error budget by running it on simulated light curves and obtained values of the order of $1\%$, which is not significant compared to typical $R_p/R_{\star}$ posterior widths.

\subsubsection{Inference process}
\label{sec:inference process}
In this work, we sample the effective parameter space using grids. This is not essential to the algorithm; we chose grids for simplicity, for ease of calibration, and to inspect the posterior across the whole parameter space. More discussion about it can be found in Section~\ref{sec:discussion}.

We evaluate the posterior on a coarse ($\log\tau$, $b$) grid ($60\times 30$, $\Delta\log\tau=0.1$) for all $\mathbf{u}$, and then on a fine grid ($800\times 30$, $\Delta\log\tau=0.0075$) restricted to the region where the coarse posterior is concentrated. For each $\mathbf{u}$ and each $b$, the fine cells are retained within the contiguous range of coarse $\log\tau$ cells around the column maximum whose posterior exceeds 1\% of that maximum, extended by one coarse cell on each side; the union of these ranges over all $\mathbf{u}$ is then evaluated on the fine grid. The threshold is applied per $b$ column, so the pre-selection reduces the cost only along $\tau$ and never discards impact-parameter values.

Both grids are uniform in $\log\tau$ (natural log) over $[-6,0]$ and non-uniform in $b$: the cells are uniform in the transformed variable $b_{\text{transf}}=a(1-(1-c\cdot b)^m)$
with $a\approx1.75$, $c\approx0.99$, $m=0.2$, chosen such that $b_{\text{transf}}(1)=1$ and the cell width shrinks by a factor of 30 from $b=0$ to $b=1$. This increases the resolution towards large $b$, where the transit model changes more rapidly with $b$.

In our algorithm, the steps are executed per target as the following sequence:
\begin{itemize}[noitemsep, topsep=0pt, left=10pt, label=-]
    \item Obtain the stellar parameters and their priors, 
    % (Section~\ref{sec:stellar_parameters})
    prepare the whitened folded binned light curve.
    % (Section~\ref{sec:target_preparation})
    \item From the distributions of $T_\star$, $\log g_\star$, $Z_\star$, we prepare the $\mathbf{u}$ grid and prior (Figure~\ref{fig:ldc_sampling}).
    % (Section~\ref{sec:limb_darkening_coefficients})
    \item From $P$ and the distributions of $R_\star$, $M_\star$, $i$, $e$, $\omega$, prepare $(\tau, b)$ priors on a coarse grid that can be interpolated.
    \item Prepare two grids of $(\tau, b)$.
    % \item Construct the templates, converting the $(\tau, b)$ grid to the $(a^e, i^e)$ grid 
    % (Section~\ref{sec:template_bank_construction}).
    \item For each $\mathbf{u}$, compute posteriors on the coarse $(\tau, b)$ grid, identify where the posterior is localized and repeat the process with a fine grid.
    % \item Iterate over $(u_1, u_2)$ and $(\tau, b)$, generate transit templates, and whiten them. Match each template to the data and compute the Gaussian posterior on $(R_p/R_\star)^2$ 
    % (Section~\ref{sec:planetary_radius}).
    \item Marginalize over the grid parameters to compute the final $(R_p/R_\star)^2$ distribution.
\end{itemize}

An example of one processed Kepler target is presented in Appendix~\ref{app:inference_process_illustration}.

\section{Results}
\label{sec:results}

\subsection{Performance of the method in simulation}
\label{sec:performance_simulation}
To validate the method, we conduct a simulation and recovery of known planetary radii. We use the noise spectrum shape of an arbitrarily selected Kepler target (KIC 002695110). We use a period of $\approx44.2$ days, so that a planet would have 39 transits during Kepler observing time. We sample the orbital parameters of the synthetic transits according to the priors described in Section~\ref{sec:template_bank_construction}. For planetary radii, we use a flat prior $(R_p/R_{\star})^2\sim U[0.01^2, 0.04^2]$. We simulate $10^4$ folded whitened transit light curves and run our inference on them.

\paragraph{p-value distributions of the true radii}
To validate our method, we consider the distributions of the p-values (the one-sided survival function) of the true radii in the posteriors reported by our algorithm~\citep[][]{cook_2006, talts_2018}. 

\begin{figure}[ht]
    \centering
    \includegraphics[width=0.47\textwidth]{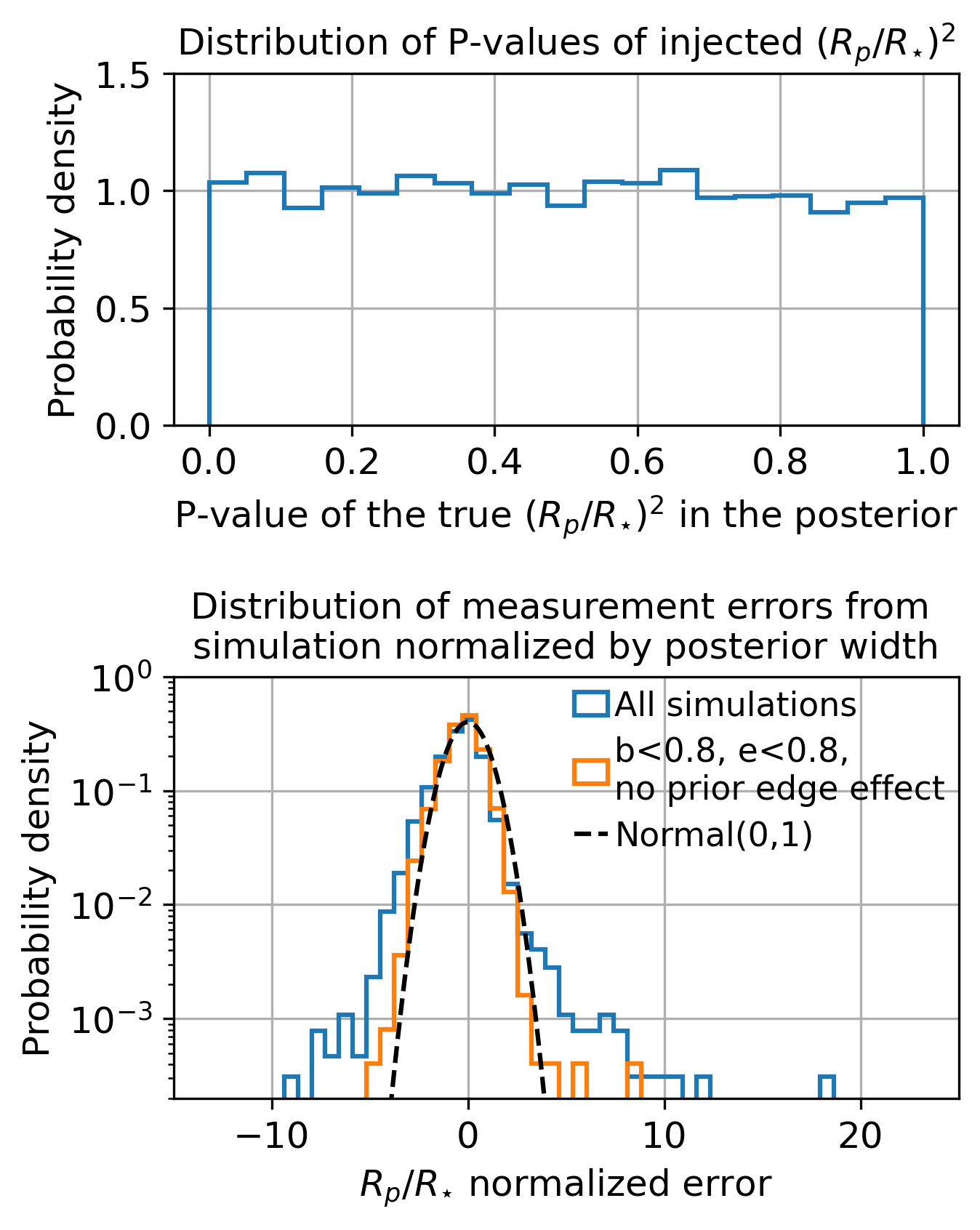}
    \caption{Upper panel: distribution of the p-values of the true simulated planetary radii with respect to the computed posteriors. As expected for a calibrated method, the distribution is close to uniform. 
    Lower panel: distribution of the difference between the best-fit value and the true value normalized by the posterior width. The blue line shows all the simulations. The orange shows cases that are not affected by the prior edge effect and have moderate eccentricities $e<0.8$ and impact parameters $b<0.8$. For comparison, the black dashed line shows the standard normal distribution that would be expected in the case of  Gaussian posteriors.}
    \label{fig:normalized_errors_distribution_and_p_vales_sim}
\end{figure}

If the true radii are sampled from the same prior as used for inference, a correct algorithm should result in a uniform distribution of these p-values. The nearly flat distribution presented in the upper panel of Figure~\ref{fig:normalized_errors_distribution_and_p_vales_sim} shows that our method is unbiased and properly captures the width of the posterior. In case of a bias, the p-value distribution would have a slope. If the posterior width were too narrow or too wide, the distribution would have excess mass in the edges or in the center. 

We note that this validation technique shows the self-consistency of the method and is valid for arbitrary posterior and prior shapes. An alternative way to characterize the performance of the method could be to look at the measurement error distribution, however its expected distribution depends on the specific shape of the posterior and the prior.

We also note that while this calibration is necessary, it is not sufficient to validate the method, as an algorithm returning the prior would pass this validation. Therefore, in Appendix~\ref{app:posterior_width_validation} we additionally compare the posterior width distribution with the theoretical expectation.

\paragraph{Measurement error distributions}
For additional illustration, we also present the distribution of the normalized measurement error, which is the deviation of the best-fit value from the truth weighted by the posterior half-width (Figure~\ref{fig:normalized_errors_distribution_and_p_vales_sim}, lower panel). For a Gaussian posterior, the distribution of these normalized errors should be unit normal. In our case, the posteriors are generally not Gaussian. One of the reasons for it is the uniform prior edge effect that truncates the posteriors, resulting in larger normalized error. This effect only appears in the calibration simulation since we had to use the same prior for truth sampling and inference, to ensure consistency and verify the uniformity of the p-values distribution. In the real-data inference, the prior edges will be far from the inferred values, and this effect will not appear.

Another case of large normalized errors appears for extreme values of eccentricity or impact parameter. Figure~\ref{fig:normalized_errors_distribution_and_p_vales_sim} additionally depicts the normalized error distribution excluding systems with large eccentricities and impact parameters or significant prior edge effect. As can be seen, in these cases the normalized error distribution is compatible with the standard normal.

\paragraph{Error budget prediction}
For an additional validation of the method calibration, we compared the reported posterior width to the analytically expected values. The results, demonstrating agreement both in median posterior width and its distribution, are shown in Appendix~\ref{app:posterior_width_validation}. 
We also repeated the expected posterior width estimation while fixing different parameters to explore their contribution to the error budget. The results presented in Appendix~\ref{app:error_budget_parameters} show that most of the error budget comes from the orbital parameters, whereas the limb darkening uncertainty has a minor contribution.

\subsection{Comparison to Kepler radii}
\label{sec:comparison}
We applied the algorithm to a set of 419 selected Kepler Objects of Interest (KOI) and compared the results to the values listed in the KOI table \citep[][]{cumulative_koi_table}. We selected all the KOIs with periods between 30 and 500 days, a planet-to-star radius ratio $<0.1$, transit duration $<1$ day, and that have stellar parameters in the Gaia catalog \citep[][]{gaia_2022_dr3}. These selections were made to match the regime of long-period low-SNR candidates and correspond to the search pipeline \citep[][]{ivashtenko_2025} whose triggers are eventually to be processed with this algorithm. Following the search pipeline, we exclude cases with a maximum single-event statistic of $>7$. We also exclude KOIs with multiple-event statistic MES$<7.75$ (MES$^2<60$), below which the KOI catalog is contaminated by false alarms \citep[][]{thompson_2018}. Additionally, we filter out KOI dispositioned as false positives. 

Since the KOI table does not provide the full posteriors but only upper and lower errors corresponding to 0.16 and 0.84 posterior quantiles, we derive the same quantile values from our posteriors. A scatter plot comparing the KOI table values and our derived values is presented in Appendix~\ref{app:koi_scatter_plot}. 

\begin{figure}[ht]
    \centering
    \includegraphics[width=0.47\textwidth]{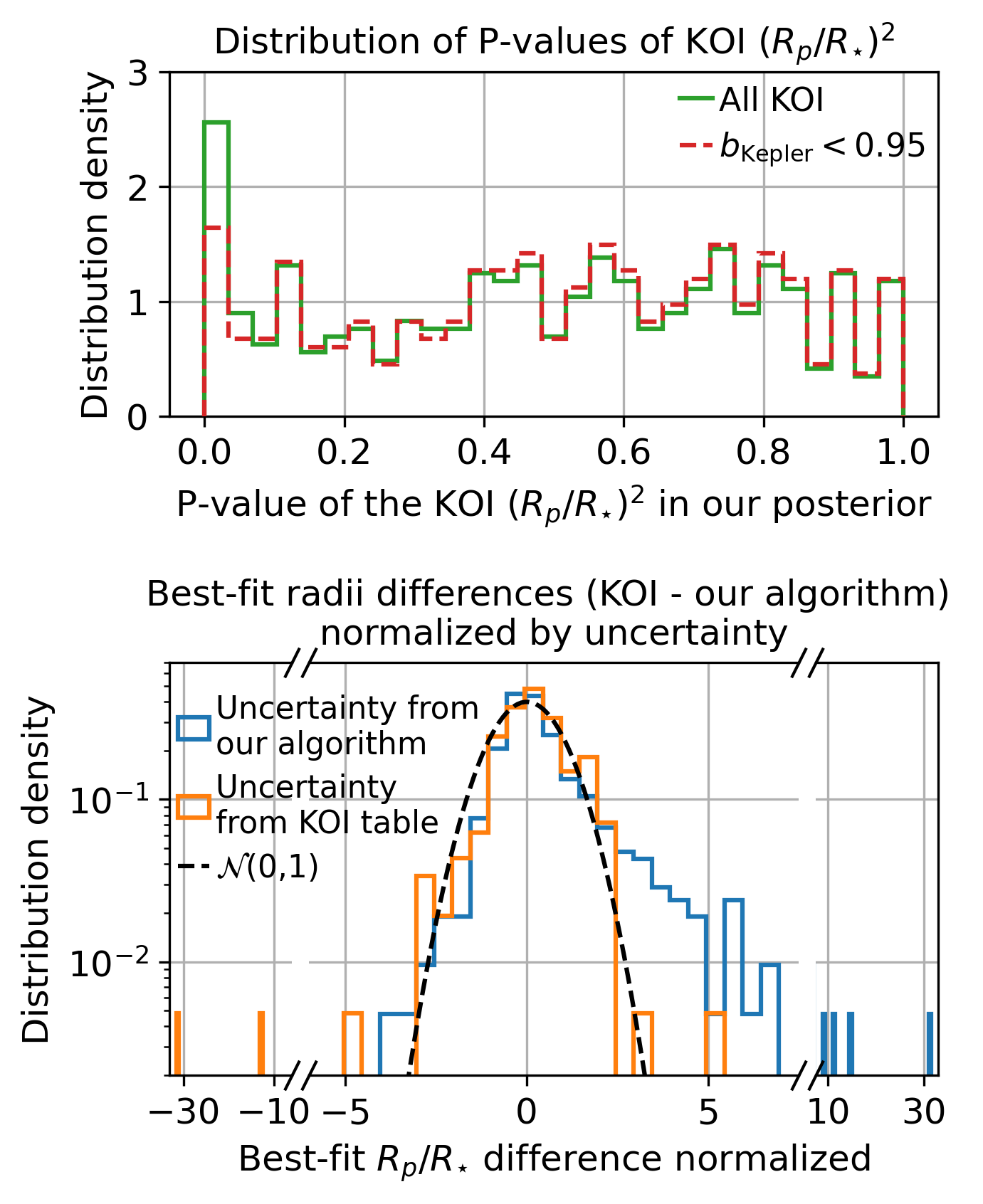}
    \caption{Upper panel: distribution of the p-values of the best fit planetary radii from the KOI table with respect to posteriors computed by our algorithm. The green line shows all the selected KOIs, while the red dashed line shows cases with KOI-table impact parameter $b<0.95$.
    Lower panel: distribution of the difference between the best-fit radius from the KOI table and from our algorithm, normalized by the uncertainty. The blue line corresponds to the uncertainty obtained from our posterior width. The orange line is for uncertainty taken from the KOI table. The black dashed line shows the standard normal distribution.}   \label{fig:normalized_errors_distribution_and_p_vales_koi}
\end{figure}

The upper panel of Figure~\ref{fig:normalized_errors_distribution_and_p_vales_koi} shows the p-values (one-sided survival function) of the KOI table best-fit radii in our posteriors. In the lower panel, we show the normalized radii difference distribution. We consider the differences between the KOI and our best-fit radii and normalize them either by the KOI or our posterior width. Two KOIs with least-squares-only fits and no reported uncertainties are excluded from the panels normalized by the KOI uncertainty.

The p-value distribution is mostly close to uniform, apart from about $5\%$ of the systems where the KOI table radius is significantly larger than our posterior predicts. As shown by the dashed line in Figure~\ref{fig:normalized_errors_distribution_and_p_vales_koi}, many of the low p-values come from the systems where the KOI-table impact parameter is $>0.95$. Due to the impact parameter-radius degeneracy, larger $b$ leads to larger estimated radii.

For some targets (e.g. KOI-4890.01), KOI table reports $b>1$ (grazing configurations, relevant for large planets), which our algorithm does not allow.
Our method therefore also truncates the corresponding high-radius tail of the posteriors. This effect is visible in the lower panel of Figure~\ref{fig:normalized_errors_distribution_and_p_vales_koi}: the excess of positive differences is smaller when normalized by the KOI uncertainties than by ours. The outliers of the normalized-error distribution are mainly cases where our and the KOI-table values of $b$ disagree. For some of them, the Kepler Data Validation values of $b$ differ from the KOI-table values as well (for example, KOI-4890.01, KOI-2535.01, KOI-1145.01).

For the most part, the normalized error spread is similar to the standard normal distribution, which is consistent with the two estimates agreeing within their stated uncertainties. However, since both analyses use the same data, their results are correlated, so the difference distribution should be narrower than the standard normal and the p-values should concentrate around 0.5 rather than be flat. The observed spread indicates that discrepancies in methods and assumptions between KOI table and our algorithm are not negligible, despite the broad agreement of the results. We regard this comparison as a consistency check, while the calibration of the method rests on the simulations of Section~\ref{sec:performance_simulation}.

We note that the NASA Exoplanet Archive \citep[][]{christiansen_2025_nasa_archive} lists several radius determinations per KOI from different works and methods, as well as the DV values \citep[][]{li_2019_dv_fit}, and these do not always agree with each other. We compare only to the KOI table, as a test of general consistency.

\section{Discussion}
\label{sec:discussion}

\subsection{Computational cost}
As a measure of computational cost, we use the number of transit model generations and likelihood computations that we need to perform per target. Quoting wall time is not meaningful, since it depends on the implementation of the method, programming language and the computing resources used. The goal of this work is to validate the concept of the method and not optimize its technical implementation. In the current implementation, not optimized for speed, our run on Kepler KOIs resulted in the median CPU time on a one-thread node of 48 minutes.

\paragraph{Reference costs}
For reference, we quote computational costs from several previous works, in which some model parameters were fixed to their assumed values. 
The Kepler MCMC procedure \citep[][]{rowe_2014_koi_validation_and_fit} generated four Markov chains of $10^6$ samples each, for each planetary candidate. It sampled 6 parameters, assuming circular orbits and fixing limb darkening coefficients to their tabulated values.
In \citet[][]{niraula_2017} and \citet[][]{del_ser_2020}, the same assumptions were used, and $\sim10^6$ samples were required.
\citet[][]{alexoudi_2022} kept the limb darkening fixed and the orbit circular but included three free time polynomial terms, requiring $6\times10^5$ samples.
In \citet[][]{maxted_2023}, the orbit was fixed to circular and a four-parameter limb-darkening law was used, requiring $\sim10^5$ samples.

\paragraph{Current costs in this work}
In this work, we sampled the effective parameters using grids. In this deliberately conservative approach, we had to make about $10^5$ transit model and likelihood computations per target. 
Our method assumes neither a circular orbit, nor exactly known stellar parameters, nor theoretical limb darkening coefficients. 
% It preserves the error budget in the $R_p/R_{\star}$ posterior from these parameters, while only sampling a reduced 4-parametric space $(\mathbf{u}, \tau, b)$. 
Thus, at equal or smaller cost, we marginalize over more parameters 
% (eccentric orbits, stellar parameters, LDC uncertainty) 
to get a more realistic $R_p/R_{\star}$ posterior.

\paragraph{Projected cost limit}
We have not optimized the grid. We used the same grid cell sizes to cover targets with different SNR, while low-SNR targets could be processed with a coarser grid. As can be seen in Figure~\ref{fig:marg_distributions_example}, we also compute likelihoods for a large fraction of grid cells with negligible posterior weight. Less conservative adaptive grids could provide an order of magnitude reduction in the number of models and likelihoods that need to be computed.

Using grids is not an essential part of the method. The $(\mathbf{u}, \tau, b)$ space can be sampled with MCMC or any other way. Such sampling will be more efficient since it will not compute likelihoods in regions of parameter space with negligible posterior mass. To estimate the required number of effective samples, we consider the $\lambda$ posterior $\alpha$-th quantile $q_{\alpha}$. We ask how many effective samples $N_{\text{eff}}$ are needed to get the estimated $\hat{q}_{\alpha}$ from the estimated posterior with a required relative precision. We use the result for the quantile variance from \citet[][Sec.~2.5.2]{serfling_1980_book},
\begin{equation}
    \text{Var}\left( \hat{q}_\alpha \right)
    \approx
    \frac{\alpha(1-\alpha)}
    {N p^2\left(q_\alpha \right)},
    \label{eq:var_quantile}
\end{equation}
where $p(q_\alpha)$ is the probability density function of the posterior at the required true quantile. Inverting this equation, we can estimate $N_{\text{eff}}=N$ for which the quantile error equals a requested value. We used a subsample of our simulations (Section~\ref{sec:performance_simulation}) and estimated $N_{\text{eff}}$ needed to get the $0.84$ quantile with relative precision of $5\%$. We obtained $N_{\text{eff}}$ of about $10^2-10^3$, depending on SNR and system parameters. 

If MCMC is used, the actual number of model and likelihood evaluations,  $N_{\text{eval}}\sim\tau_{\text{int}}N_{\text{eff}}$ is larger due to the integrated autocorrelation time $\tau_{\text{int}}$. Typical values $\tau_{\text{int}}\sim10^0-10^2$ depend on the sampler.   

\subsection{Improvement directions of the current method}
This method extends directly to the posterior of the planetary radius in absolute units, which the occurrence-rate inference ultimately needs. Appendix~\ref{app:rp_abs_units_posterior} shows that this requires modifying only the prior computation and the marginalization equation, with no additional likelihood evaluations.

The specific implementation of this method using grids also allows, in principle, computing the evidence integral. It can be useful in calculating the evidence ratio distinguishing planetary signals from false positives and false alarms~\citep[][]{matesic_2024}.

The dimensionality reduction used in this work can be taken further. Here, we used analytical dimensionality reduction based on the approximate equations for transit duration and impact parameter. These two parameters are strongly correlated, meaning that a more efficient parametrization can be found.

For example, following the idea of \citet[][]{roulet_2019_template_bank}, one can use principal component analysis (PCA) to derive numerically the basis of non-physical templates reproducing any transit model in the most efficient way. In \citet[][]{wadekar_2023}, this approach was taken even further, combining analytical tools and machine learning. This allowed dimensionality reduction for more general cases when a lower-dimensional manifold is embedded into a higher-dimensional degenerate parameter space. Such a case can be a bottleneck for affine-invariant ensemble samplers that handle linear degeneracies well but not curved ones.

Beyond transit light curves, the same idea can be applied to the microlensing signals which also suffer from degeneracies when multiple parameter combinations lead to the same model. Similarly to transits, the shape of a microlensing model is relatively simple, and it should be possible to parametrize it with a smaller parameter basis.

\section{Conclusion}
\label{sec:conclusion}

We presented a method for computing marginalized posterior distributions of the planet-to-star radius ratio for low-SNR transiting planets. The method reduces the 11-parameter transit inference problem to an effective 4-parameter space $(\tau, b, \mathbf{u})$. The likelihood in $\lambda=(R_p/R_{\star})^2$ is computed analytically, the transit epoch is marginalized using the convolution theorem, and the remaining orbital and stellar parameters are compressed into template priors that preserve their probability mass. The resulting posterior therefore retains the error budget of all the nuisance parameters, including eccentric orbits and stellar parameter uncertainties, while the transit model and likelihood are computed only for the templates. The method operates on whitened folded light curves, mitigating biases from the correlated noise.

We validated the method on $10^4$ simulated systems, obtaining a uniform distribution of the p-values of the true radii and posterior widths that agree with an independent semi-analytical prediction. Most of the error budget comes from the poorly constrained orbital parameters (under our broad eccentricity prior), while the limb darkening uncertainty is subdominant. Applied to 419 faint long-period Kepler Objects of Interest, the method yields radius ratios broadly consistent with the KOI table values, however the spread between the two methods is broader than expected.

The derived posteriors are intended for the hierarchical occurrence rate inference of small long-period Kepler planets detected by our search pipeline \citep{ivashtenko_2025}, where each candidate contributes its full radius posterior rather than a point estimate. The same dimensionality-reduction approach can be extended with numerically constructed template spaces. It can be applied to other survey-scale inference problems, such as PLATO and Roman transit candidates, and can be adapted to microlensing events.

\section*{Acknowledgements}

This research was generously supported by the Shimon and Golde Picker - Weizmann Annual Grant, the Israeli Science Foundation (ISF) and the Israeli Council for Higher Education (CHE) via the Weizmann Data Science Research Center, and by a research grant from the Estate of Harry Schutzman.

This research has made use of the NASA Exoplanet Archive~\citep[][]{akeson_2013_nasa_archive}, which is operated by the California Institute of Technology, under contract with the National Aeronautics and Space Administration under the Exoplanet Exploration Program.

\facility{Kepler}

\software{
\texttt{numpy} \citep{numpy},
\texttt{scipy} \citep{scipy},
\texttt{batman} \citep{kreidberg_2015_batman},
\texttt{matplotlib} \citep{matplotlib},
\texttt{Jupyter} \citep{jupyter}
}

AI tools were used to gather references, assist in writing auxiliary parts of code and debugging, suggest language improvements within the manuscript, and proofread. The AI tools used were Claude and ChatGPT.

The posteriors for the 419 KOIs analyzed in Section~\ref{sec:comparison} are available from the authors upon request.

\appendix

\section{Inference process illustration}
\label{app:inference_process_illustration}

\begin{figure*}[]
    \centering
    \includegraphics[width=0.75\textwidth]{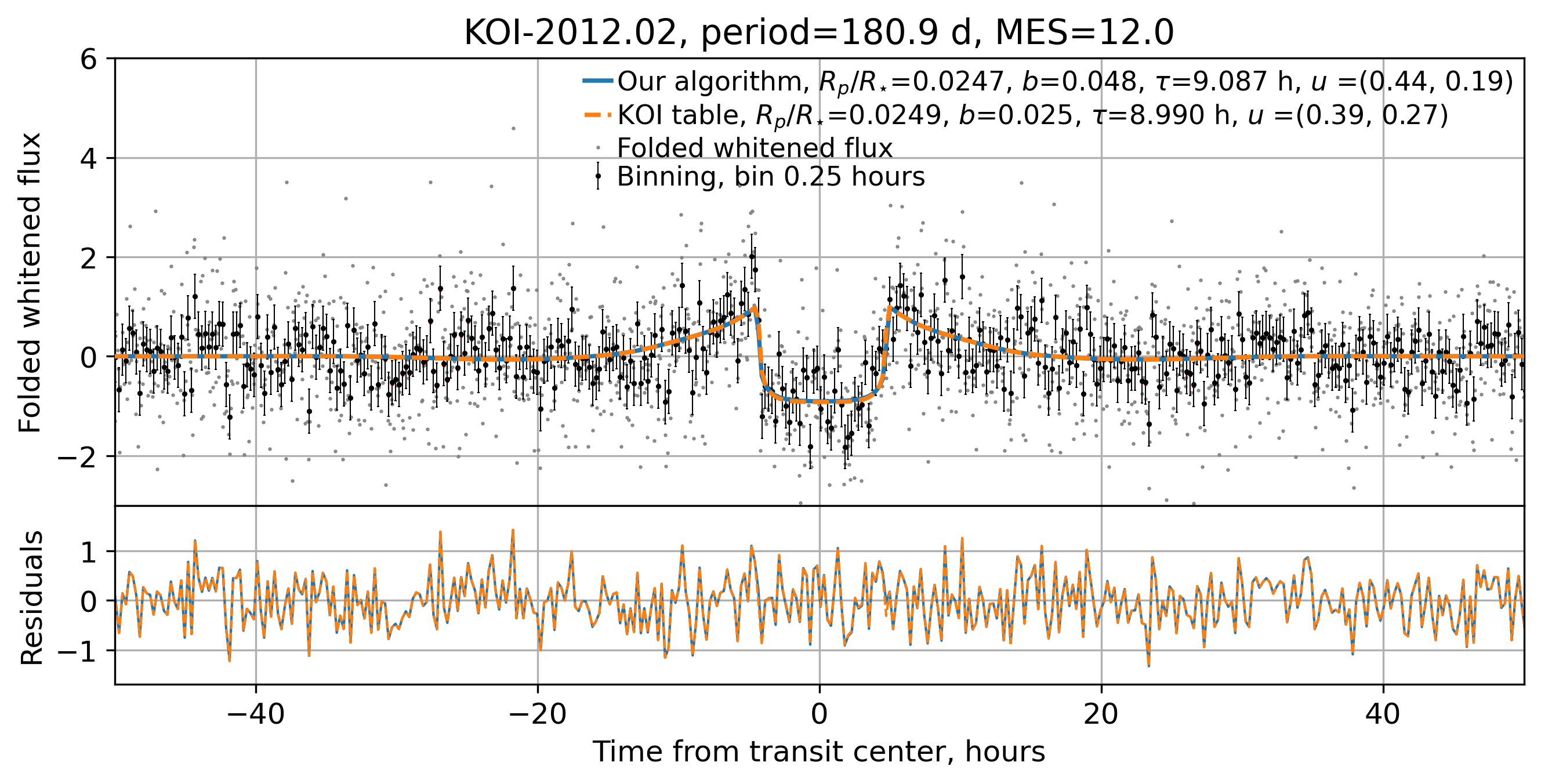}
    \caption{Upper panel. Gray dots: folded whitened flux values for KOI-2012.02. Black dots with error bars: whitened folded binned flux. The error bars correspond to the binned standard deviation. The flux is prepared as described in Section~\ref{sec:target_preparation}.
    Blue line: maximum a posteriori (MAP) fit from our algorithm, with the model generated as a pseudo-circular orbit corresponding to the specified duration and impact parameter, as described in Section~\ref{sec:template_bank_construction}. We note that we do not assume the orbit to be circular: a similar-looking transit may correspond to non-zero eccentricity. Orange line: the model with parameters listed in the KOI table \citep[][]{ cumulative_koi_table}.
    Lower panel: residuals for the two models.
    }
    \label{fig:folded_fux_fit_koi_example}
\end{figure*}

This appendix provides an illustration of the algorithm operation on an example of Kepler Object of Interest KOI-2012.02, a confirmed planet (Kepler-1052 c) with a period of 180.92 days, radius $2.3^{+0.5}_{-0.2} R_\oplus$, radius ratio $0.0249^{+0.0013}_{-0.0008}$, and maximum multiple event statistic (MES) of 12.03 \citep[][]{valizadegan_2023, cumulative_koi_table}. Figure~\ref{fig:folded_fux_fit_koi_example} shows its whitened folded binned flux, prepared as described in Section~\ref{sec:target_preparation}. It also plots the best-fit template from our algorithm, compared with the solution from the KOI table \citep[][]{ cumulative_koi_table}, showing agreement in this case. 

\begin{figure*}[]
    \centering
    \includegraphics[width=0.75\textwidth]{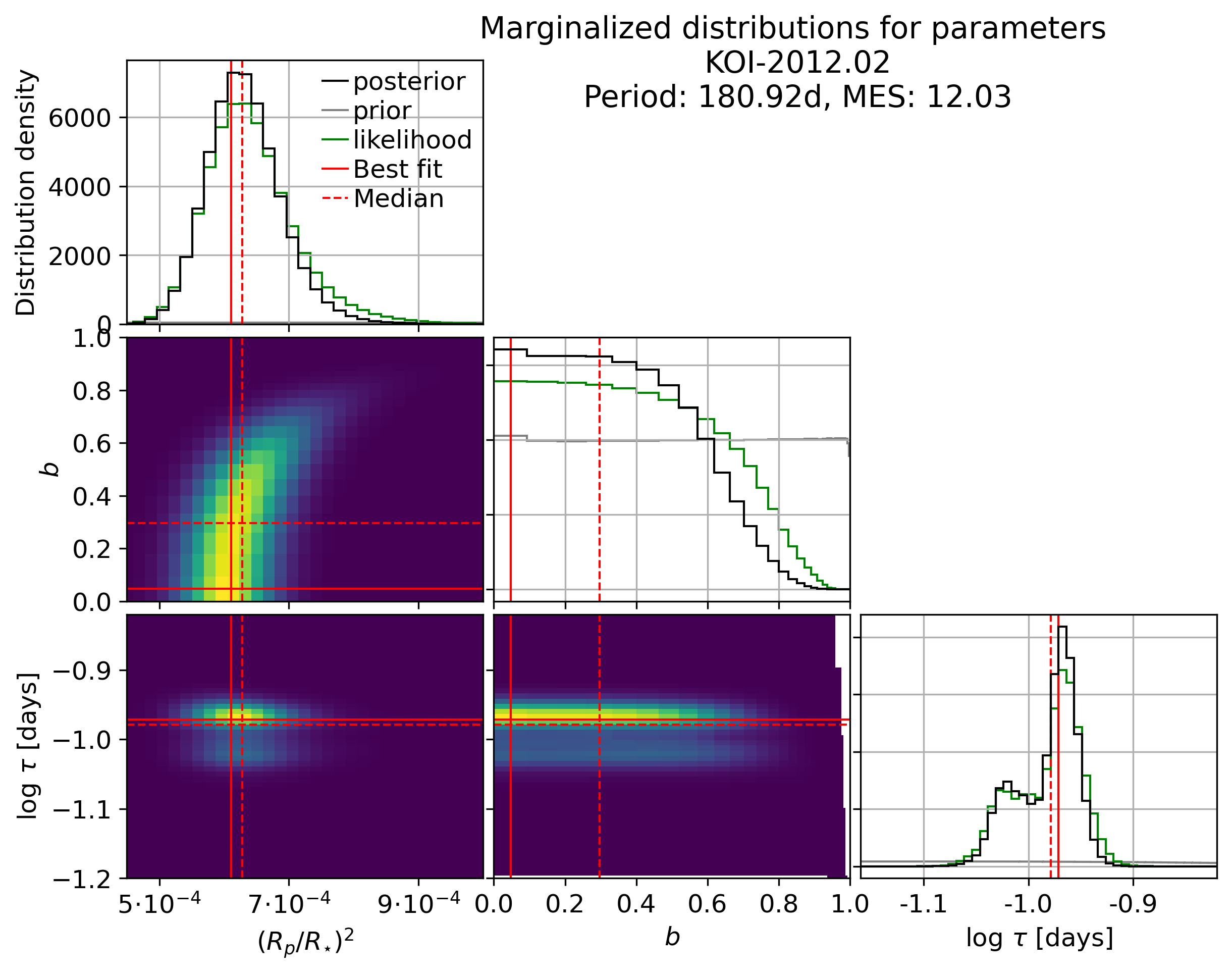}
    \caption{Corner plot of posterior distribution of $(R_p/R_\star)^2$ and effective impact parameter $b$ and natural logarithm of transit duration $\tau$ of the templates used for inference. Example for one of the targets, KOI-2012.02. The distributions are marginalized over the limb darkening parameters used for the inference.}
    \label{fig:marg_distributions_example}
\end{figure*}

Figure~\ref{fig:marg_distributions_example} shows the outputs of our algorithm for KOI-2012.02. It provides the $(R_p/R_\star)^2$ posterior calculated as described in Section~\ref{sec:planetary_radius} on a grid. It also shows the posterior distributions of effective transit duration $\tau$ and impact parameter $b$, defined in Section~\ref{sec:template_bank_construction}.
These are not the physical duration and impact parameter of a full orbital fit, but the effective template bank values describing the light-curve shape (Section~\ref{sec:template_bank_construction}). The $(R_p/R_\star)^2$ posterior, in contrast, is physical: it includes the spread over all physical configurations consistent with the data, weighted by their priors.

The $(R_p/R_\star)^2$ posterior is mildly non-Gaussian, with the skewness depending on the SNR and system parameters. The $b$ distribution is strongly non-Gaussian and influenced by the prior, and $b$ itself is not well-constrained. The $b$ vs $(R_p/R_\star)^2$ plot shows the degeneracy between them: for larger $b$, larger radii are needed to explain the observed transit depth. This degeneracy, together with poorly measured $b$, gives one of the dominant contributions to the total $(R_p/R_\star)^2$ uncertainty budget.

The $\tau$ posterior shows a bimodal distribution for this target, which is caused by the noise realization of this dataset rather than by a physical feature.

% ----------------------------------------------------
\section{Kepler Objects of Interest scatter plot}
\label{app:koi_scatter_plot}
\begin{figure*}[]
    \centering
    \includegraphics[width=0.7\textwidth]{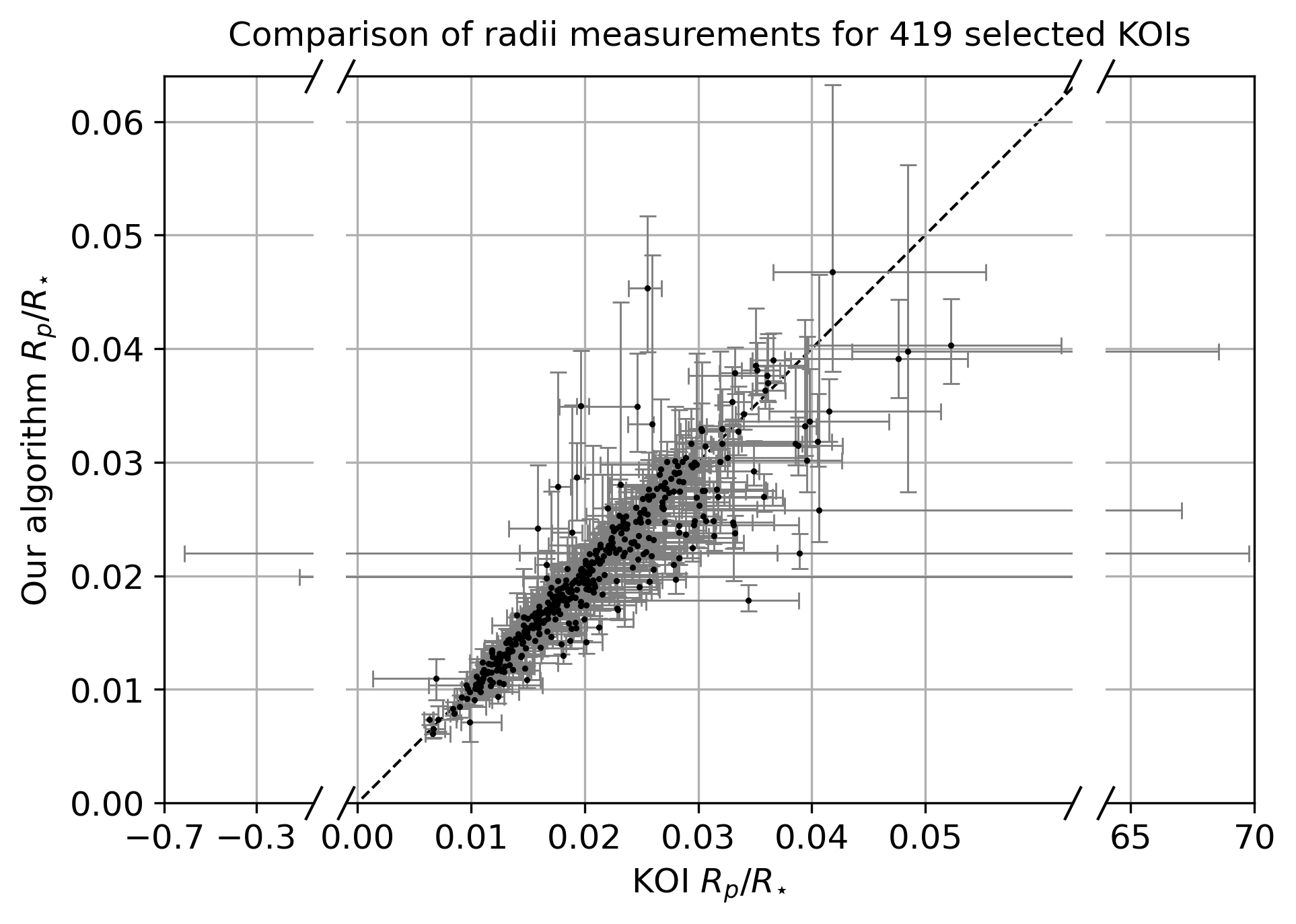}
    \vskip -0.3cm    
    \caption{Comparison of $R_p/R_\star$ and its uncertainty inferred by our algorithm versus the KOI table values \citep[][]{cumulative_koi_table}.}
    \label{fig:comparison_sctter_koi}
\end{figure*}

Figure~\ref{fig:comparison_sctter_koi} shows the inferred $R_p/R_\star$ values for all the investigated KOI from Section~\ref{sec:comparison}. It compares the $R_p/R_\star$ and its uncertainty inferred by our algorithm with the values listed in the KOI table \citep[][]{cumulative_koi_table}.

% ---------------------------------------------------------------------
\section{Planetary radius in absolute units}
\label{app:rp_abs_units_posterior}

Occurrence rate inference requires the posterior of the planetary radius $R_p = \sqrt{\lambda}R_\star$ in absolute units, rather than of the radius ratio derived in this work. Since $R_\star$ enters the inference through the prior $\pi(a/R_\star)$ (Section~\ref{sec:stellar_parameters}), the posteriors of $\lambda$ and $R_\star$ are not independent, and the correct conversion requires their joint posterior. Below we show that it can be obtained in the framework of our algorithm by modifying only the prior computation.

\paragraph{Joint posterior}
Repeating Equation~\ref{eq:summary_equation_initial} but marginalizing over all the nuisance parameters except $R_\star$ yields
\begin{align}
\begin{split}
    p(\lambda, R_{\star}|\mathbf{d})
    \approx
    \frac{\pi(\lambda)}
    {\mathcal{L}(\mathbf{d})}
    &\sum_k \Bigg[
    \pi(\boldsymbol{\eta}_k)
    \pi(R_{\star} | \boldsymbol{\eta}_k)
    \times
    \\& \times
    \mathcal{L}_{\hat{\lambda}}(\mathbf{d}|\boldsymbol{\eta}_k)
    \exp{\left(\!\!-\frac{(\lambda\!-\!\hat{\lambda}_k)^2}{2\sigma^2_{\lambda,k}}\!\right)}
    \Bigg],
    \label{eq:summary_equation_r_star}
\end{split}
\end{align}
where $\pi(R_{\star}|\boldsymbol{\eta}_k)=\pi(R_\star,\boldsymbol{\eta}_k)/\pi(\boldsymbol{\eta}_k)$ is the prior distribution of $R_\star$ conditioned on the template, and the joint prior $\pi(R_\star,\boldsymbol{\eta}_k)$ is the analog of Equation~\ref{eq:template_prior} with the integration restricted to a given $R_\star$. 

Thus, the marginal $\lambda$ posterior is unchanged, and the only new ingredient is $\pi(R_{\star}|\boldsymbol{\eta}_k)$. It is obtained from the same Monte Carlo sample that produces the template prior (Section~\ref{sec:template_bank_construction}): while accumulating the prior weights of the $(\log\tau, b)$ cells, one additionally records the $R_\star$ values of the samples falling into each cell. 

Similarly to $R_\star$, other nuisance parameters could be isolated to get their distributions.

\paragraph{Absolute radius posterior}
In the same way, we can modify Equation~\ref{eq:survival_function_no_prior} to get the survival function of $R_p=\sqrt{\lambda}R_\star$ for the improper flat prior in $\lambda$,
\begin{align}
\begin{split}
    S_0(R_p)
    = \frac{1}{\sum_k w_k} 
    &\sum_k w_k \Bigg[
    \int\! dR_\star\, \pi(R_{\star}|\boldsymbol{\eta}_k)\,
    \times
    \\&\times
    S_G\!\left(\!\left(\frac{R_p}{R_{\star}}\right)^2; 
    \hat{\lambda}_k, \sigma_{\lambda,k} \!\right)\!
    \Bigg]
    \label{eq:rp_survival_function}
\end{split}
\end{align}
with the weights $w_k$ from Equation~\ref{eq:survival_function_summation_weights}. This is the analog of $S_0(\lambda)$ with the survival function of each template evaluated at the radius ratio implied by $R_\star$. 

If a prior on $\pi(R_p)$ needs to be added, replacing the flat prior on $\lambda$, the survival function of each template is evaluated with this prior, giving
\begin{widetext}
\begin{align}
\begin{split}
    S_\pi(R_p)
    = \frac{
    \sum_k w_k
    \int\! dR_\star\, \pi(R_{\star}|\boldsymbol{\eta}_k)
    \int_{R_p}^{\infty}\! dr\, \pi(r)\,
    \mathcal{N}\!\left( (r/R_\star)^2; \hat{\lambda}_k, \sigma_{\lambda,k} \right)
    }{
    \sum_k w_k
    \int\! dR_\star\, \pi(R_{\star}|\boldsymbol{\eta}_k)
    \int_{0}^{\infty}\! dr\, \pi(r)\,
    \mathcal{N}\!\left( (r/R_\star)^2; \hat{\lambda}_k, \sigma_{\lambda,k} \right)
    },
    \label{eq:rp_survival_function_with_prior}
\end{split}
\end{align}
\end{widetext}
where $\mathcal{N}$ is the Gaussian probability density function. 

In practice, Equations~\ref{eq:rp_survival_function} and \ref{eq:rp_survival_function_with_prior} can be evaluated by sampling: a template $k$ is drawn with probability proportional to $w_k$, $R_\star$ from the prior samples that fell into this template, and $\lambda$ from the corresponding distribution; the $R_p$ quantiles are obtained from the resulting sample.

\paragraph{Remarks}
Since the relative $\lambda$ posterior width decreases with increasing SNR, the $R_{\star}$ uncertainty of about $\gtrsim6\%$ (Section~\ref{sec:stellar_parameters}) can dominate the $R_p$ error budget for high SNR (Appendix~\ref{app:error_budget_parameters}). 

Finally, we note that the $R_p$ posterior obtained in this way is conditioned on the interim prior on $\lambda$ or $R_p$ used for inference. For the hierarchical population inference, this interim prior has to be divided out when the population model is introduced \citep[][]{foreman_mackey_2014_population_inference, mandel_2019_gw_distributions}. 
% We therefore recommend delivering joint posterior samples of $(\lambda, R_\star)$ together with the interim prior, rather than the summary quantiles of $R_p$ alone.

% Neglecting the correlation, that is replacing $\pi(R_{\star}|\boldsymbol{\eta}_k)$ with $\pi(R_{\star})$, reduces Equation~\ref{eq:rp_survival_function} to the product of independent variables, $\log R_p = \log\sqrt{\lambda} + \log R_\star$, with the $\lambda$ posterior of this work and the stellar prior. This is accurate when the data do not inform $R_\star$, which is expected in the low-SNR regime: the light curve constrains $R_\star$ only through the stellar density, via the transit duration, and this constraint is degenerate with the poorly constrained eccentricity and impact parameter. The approximation can be verified per target by comparing $p(R_\star|\mathbf{d})=\sum_k p(\boldsymbol{\eta}_k|\mathbf{d})\pi(R_{\star}|\boldsymbol{\eta}_k)$ with $\pi(R_\star)$. 

% ---------------------------------------------------------------------

\section{Posterior width distribution validation}
\label{app:posterior_width_validation}

The output of our method for a single system is the marginalized posterior of $R_p/R_\star$, and here we validate its width, which is half the distance between the 0.16 and 0.84 quantiles. This width is a random variable depending on the noise realization of the data, so we examine its distribution. 
Using the simulated systems from Section~\ref{sec:performance_simulation}, we compare the distribution of the $R_p/R_\star$ posterior widths reported by our method with a theoretical prediction. The theoretical prediction (outlined below) uses the same priors and the same transit model as our pipeline, but involves no template bank, no low-dimensional reparameterization, no linearization in $\lambda$, no representation of the $\lambda$ posterior as a Gaussian mixture, and no common code with the pipeline. It thus additionally validates the assumptions made in our pipeline and ensures it introduces no significant biases.

\begin{figure}[]
    \centering
    \includegraphics[width=0.47\textwidth]{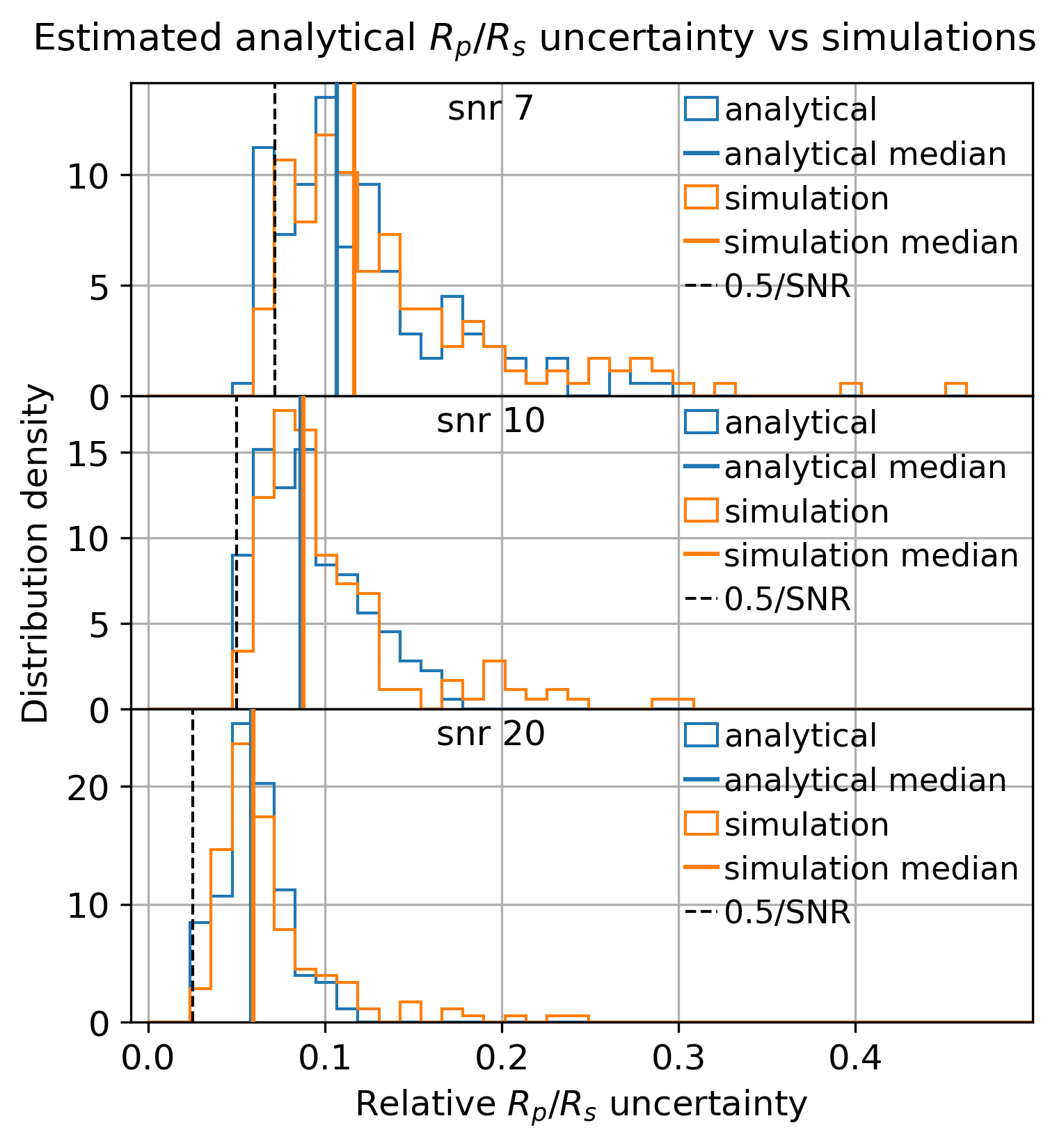}
    \caption{Distributions of relative $R_p/R_\star$ posterior widths, or posterior widths normalized by the median value. The three panels correspond to different SNR bins. Blue line shows the distributions obtained from theoretical estimation. Orange line stands for the values obtained with our algorithm. Black dashed line is the uncertainty floor when all the nuisance parameters are fixed to their true values.}
    \label{fig:analyt_vs_simulated_uncert_histograms}
\end{figure}

In Figure~\ref{fig:analyt_vs_simulated_uncert_histograms}, we compare the distributions of the relative posterior widths obtained from our pipeline run (Section~\ref{sec:performance_simulation}) and from the theoretical evaluation. We bin the result by SNR and present the distribution for multiple systems with different parameters sharing the same SNR bin. All the distributions show a general agreement between our pipeline and the theoretical expectation. 

We also plot the median uncertainty values and the reference lower thresholds of $0.5/\text{SNR}$. This lower threshold is the uncertainty expected if all the nuisance parameters are fixed to their true values. As can be seen, there is a significant difference between the actual medians and this lower threshold, arising from the uncertainties in the nuisance parameters. Appendix~\ref{app:error_budget_parameters} further explores the uncertainty budget of the nuisance parameters.

The theoretical prediction for the $R_p/R_\star$ posterior width and its distribution arises from evaluating the integral over the nuisance parameters in Equation~\ref{eq:summary_equation_initial}. The integral in Equation~\ref{eq:summary_equation_initial} has to be evaluated on a grid of $R_p/R_\star$ and over the noise realizations, which drive the variation of the posterior width.

We evaluate the 6-dimensional integral over $\boldsymbol{\theta}$ by Monte Carlo over the prior, using importance sampling. Computing the full integral brute-force with loops over $R_p/R_\star$ and noise realizations would be computationally heavy, so we use two simplifications. 

First, we use the fact that the model $\mathbf{s}$ is a smooth function of $R_p/R_\star$, comprising $(R_p/R_\star)^2$ depth scaling and slow changes in the transit shape. We
represent the model as a linear combination of a few basis vectors, precompute the basis inner products, and evaluate the likelihood thereafter without model calls \citep[][]{field_2014, trefethen_2019}.

Second, we use the fact that data enters the likelihood linearly and only through projections, allowing us to reduce the expression to a sufficient statistic \citep[][]{bretthorst_1998}. Thus, we never have to generate data with noise; the noise enters as a random shift of the noiseless sufficient statistic.

We do not use the Fisher information matrix, the standard tool for forecasting parameter uncertainties, because it assumes a Gaussian posterior, which does not hold here: there are curved degeneracies, and at low SNR the nuisance parameters are constrained by their priors. In addition, the Fisher forecast is deterministic: it attaches one number to the noiseless signal and gives no information about posterior width variation between noise realizations.

\section{Error budget}
\label{app:error_budget_parameters}
\begin{figure}[]
    \centering
    \includegraphics[width=0.47\textwidth]{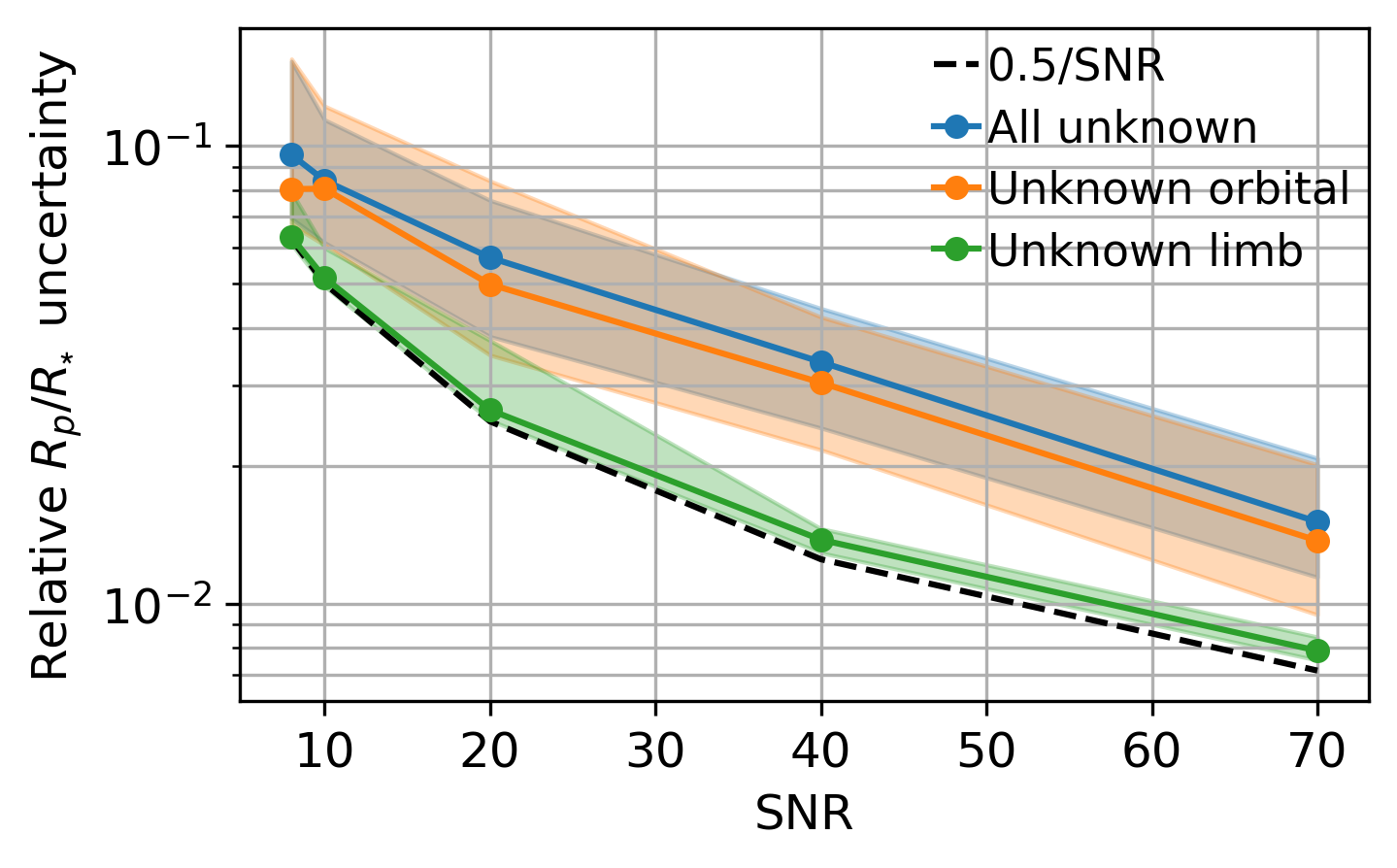}
    \caption{Contributions of orbital parameters ($e$, $i$, $\omega$, $a$) and limb darkening parameters ($u_1$, $u_2$) to the $R_p/R_\star$ posterior width, as a function of SNR. The blue line shows the expected posterior width for simulated transit models from Section~\ref{sec:performance_simulation}. The shaded bands show the variation of the posterior width for different systems in the same SNR bin. The orange line corresponds to the case when ($u_1$, $u_2$) are pinned to their true values. The green line is the case when ($e$, $i$, $\omega$, $a$) are pinned to their true values. The dashed black line is the uncertainty floor when all the nuisance parameters are fixed to their true values.}
    \label{fig:error_budget_analytical}
\end{figure}

This appendix investigates the contribution of the uncertainty of different groups of nuisance parameters into the $R_p/R_\star$ posterior width. 
We repeated the calculation predicting the $R_p/R_\star$ posterior width from Appendix~\ref{app:posterior_width_validation} replacing the prior of a group of nuisance parameters by a delta function at the true values. In this way, we isolate the contribution only from the parameters that are not fixed and are still integrated over their full priors. With all nuisance parameters fixed, the computation reproduces the single-parameter floor $\sigma[R_p/R_\star]/(R_p/R_\star) = 0.5/\text{SNR}$ set by the matched-filter SNR. The relative contribution of this floor to the total uncertainty budget grows with decreasing SNR, as can be seen in Figure~\ref{fig:error_budget_analytical}.

Figure~\ref{fig:error_budget_analytical} shows the expected $R_p/R_\star$ posterior width as a function of SNR for 3 cases. The first one is the general case when all the parameters ($e$, $i$, $\omega$, $a$, $u_1$, $u_2$) are integrated over their priors. The second case is when the orbital parameters ($e$, $i$, $\omega$, $a$) are integrated and the limb darkening parameters ($u_1$, $u_2$) are pinned to their true values. The third case is the opposite, when the limb darkening parameters are integrated. As can be seen, under the eccentricity prior we use, most of the error budget comes from the unknown orbital parameters, whereas the limb-darkening parameter contribution is subdominant. 

We note that this plot only shows posterior width and does not show a bias that may occur if parameters are pinned to wrong values.

\section{Assumptions of the algorithm}
\label{app:assumptions}

The method is tailored to marginalized radius posteriors of small, long-period, low-SNR planets from our search \citep[][]{ivashtenko_2025}. For this problem, we make the following simplifying assumptions:
\begin{itemize}[noitemsep, label=-, topsep=0pt]%, leftmargin=*]
    \item We assume there is no transit timing, duration, or depth variations. We take the period as exactly defined from the search.
    \item We assume the transits to be symmetric in time. This is not true for high-eccentricity transits with some periastron angles.
    \item We assume the impact parameter to be a valid description for a transit, meaning that the sky-projected planet trajectory is straight and symmetric. This assumption also fails for extreme values of eccentricity.
    \item We assume that star-planet distance in transit significantly exceeds the stellar radius. It is true for the planetary periods that we consider ($>30$ days) and dwarf stars. 
    \item We assume that the planet-to-star radius ratio is sufficiently small that the impact parameter remains $<1$, and that the planetary radius affects the transit light curve primarily through the transit depth, with a smaller effect on the ingress and egress. 
    \item We condition the analysis on the assumption that the signal is a planet and not an eclipsing binary.
    \item We neglect any light emitted by the planet. 
    \item We do not infer stellar parameters from the light curves. We use external stellar parameters and neglect correlations between them. 
    \item We neglect blending effects, assuming that the flux is due to the target star, relying on the correction of the Kepler light curves for crowding.
    \item We do not model possible spot crossings or other contaminants. We also assume that the bad cadences were removed in the search or post-processing, and all the signal considered here is due to transits.
    \item We do not explore the posterior dependence on the preprocessing and whitening procedure.
    \item We limit the analysis to the parameters introduced in 
    Section~\ref{sec:summary_method} and do not consider other effects, such as, for example, gravity darkening.
\end{itemize}

Some of these assumptions are linked to our parametrization not describing adequately cases of extreme eccentricities, as is demonstrated in Appendix \ref{app:tb_errors}. We assume these cases to be rare and not to contribute significantly to the analysis. Other assumptions are linked to the fact that we work with low-SNR signals, therefore many effects are not well-measurable.

\section{Template bank error}
\label{app:tb_errors}
In this Appendix, we verify the precision of the effective orbital parameters approximation and the template bank approach (Section~\ref{sec:template_bank_construction}). For this, we consider the amplitude of the expected transit models measured with the approximate templates. The model amplitude is defined as the estimate (Equation~\ref{eq:amplitude_estimator}) with $\mathbf{d}=\mathbf{s}_{\text{true}}$ and $\mathbf{h}=\mathbf{s}_{\text{bank}}$, where $\mathbf{s}$ is a transit model (Equation~\ref{eq:data_model}). Amplitude error is the difference between the amplitude for a given template and a perfect match amplitude. A perfect match would give an amplitude of 1 and error of 0. An error of 1 means that the measured radius squared is twice the true one.  

We generate noiseless transit models using the orbital parameter prior from Section~\ref{sec:performance_simulation}. From the parameters of each model, we calculate analytical effective orbital parameters (Section~\ref{sec:template_bank_construction}) and find the corresponding template in the template bank. We then use Equation~\ref{eq:amplitude_estimator} to measure the transit amplitude using this template.

\begin{figure*}[]
    \centering
    \includegraphics[width=0.85\textwidth]{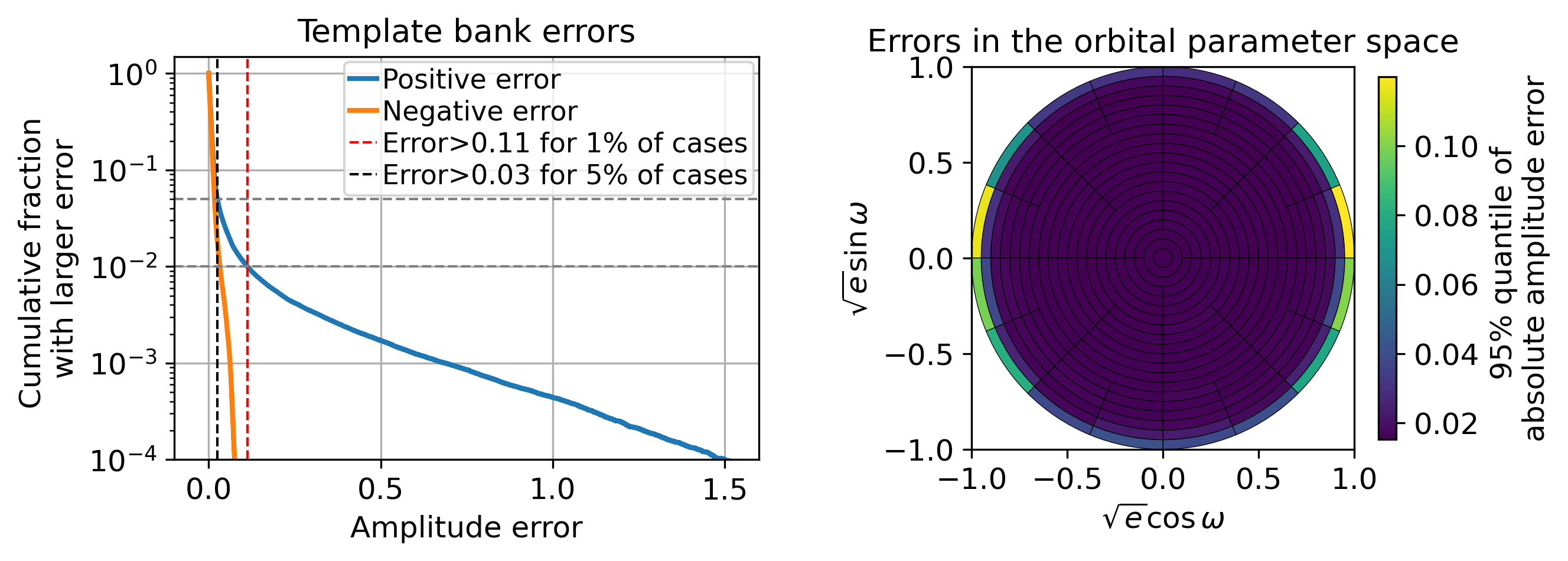}
    \caption{Distribution of errors for effective templates (Section~\ref{sec:template_bank_construction}) measuring amplitudes of transit models using Equation~\ref{eq:amplitude_estimator}. Left panel shows how frequently we get errors larger than a given one. Blue line is for positive errors, and orange line is for negative errors. The right panel considers such distributions for cells in the orbital parameter space and reports the $95\%$ quantiles for each cell.}
    \label{fig:tb_errors}
\end{figure*}

In the left panel of Figure~\ref{fig:tb_errors}, we show the distribution of the amplitude errors when our template bank method is applied to the sampled transit population. We plot a survival distribution showing how frequently we get an error larger than a given value. As can be seen, for 95$\%$ of the cases, we get errors $<3\%$, which corresponds to $<2\%$ error on the radius. In rare cases, we can get significant errors, measuring a few times larger amplitudes. The right panel of Figure~\ref{fig:tb_errors} shows that these cases correspond to large eccentricities.

In the right panel of Figure~\ref{fig:tb_errors}, we show the $95\%$ quantile of the error distribution as a function of orbital coordinates, where $e$ is eccentricity, and $\omega$ is periastron argument. For every cell in the diagram, there is a distribution of amplitude measurement errors obtained for the transits from this cell. We report the value of the error corresponding to the $95\%$ quantile of this distribution. As can be seen, all the values of $>5\%$ are located at extreme eccentricities, $e>0.9$. The $\omega$ values of the large-error cases also correspond to the asymmetric-transit configurations excluded by the symmetry assumption in Appendix~\ref{app:assumptions}.

\section{Marginalization over the transit epoch}
\label{app:t0_marginalization}

We show the modifications to the planetary radius estimator (Equation~\ref{eq:amplitude_estimator}) that account for the marginalization over the transit epoch $t_0$. We consider here the simplified case when folded flux has the same variances in different bins.

Consider a transit likelihood with two parameters, $\mathcal{L}(\mathbf{d}|\lambda, t_0)$. Using Bayes relation, marginalizing over $t_0$, and factorizing similarly to Equation~\ref{eq:factorized_likelihood_lambda_hat}, the distribution for $\lambda$ yields
\begin{align}
\begin{split}
    p(\lambda|\mathbf{d}) 
    &= \int dt_0 
    \frac{\mathcal{L}(\mathbf{d}|\lambda,t_0) \pi(\lambda) \pi(t_0)}
    {\mathcal{L}(\mathbf{d})} 
    \\&
    \propto \int dt_0 \mathcal{L}(\mathbf{d}|\lambda,t_0)
    \\&
    \propto 
    \int dt_0
    w(t_0)
    \exp\left[
    -\frac{1}{2}
    \frac{ \left(\lambda - \hat{\lambda}\right)^2}
    {\sigma_\lambda^2}
    \right],
    \label{eq:bayes_likelihood_ampl_marginalization}
\end{split}
\end{align}
where $\hat{\lambda}$ is given by Equation~\ref{eq:amplitude_estimator}, $\sigma_\lambda$ is from Equation~\ref{eq:sigma_amplitude_estimator} and is a constant. The weight factor is 
\begin{equation}
    w(t_0) \propto \exp\left[
    \frac{1}{2}\frac{\left<\mathbf{d}, \mathbf{h}\right>^2}{\left<\mathbf{h}, \mathbf{h}\right>}
    \right].
\end{equation}
This allows viewing Equation~\ref{eq:bayes_likelihood_ampl_marginalization} as a mixture of Gaussian distributions for $\hat{\lambda}$ weighted by $w(t_0)$.
Marginalizing the Gaussian mixture in Equation~\ref{eq:bayes_likelihood_ampl_marginalization} over $t_0$ results in a distribution with a variance
\begin{align}
    \sigma_{\text{tot}}^2 &= \int w(t_0) \left[
    \sigma_\lambda^2(t_0) + (\hat{\lambda}(t_0) - \mu_\lambda)^2
    \right] dt_0
    \\&=\sigma_\lambda^2 + \int w(t_0) (\hat{\lambda}(t_0) - \mu_\lambda)^2 dt_0,
    \label{eq:sigma_ampl_after_marginalizing_t0}
\end{align}
where 
\begin{align}
    \mu_\lambda = \int dt_0\; w(t_0) \hat{\lambda}(t_0).
    \label{eq:mean_amplitude_marginalized}
\end{align}
We approximate the mixture by a single Gaussian centered at the maximum-likelihood amplitude with variance $\sigma_{\text{tot}}^2$. In practice, $w(t_0)$ is evaluated in a window around the maximum-likelihood epoch and the epoch-scatter term is added in quadrature to $\sigma_{\lambda}^2$ (Section~\ref{sec:math_planetary_radius}). 

The approximation holds when the mixture is unimodal and can be described by a Gaussian. This is true for the SNR above the detection threshold, where the weights are concentrated only on one or a few shifts. 
At lower SNR ($<5$), the mixture becomes skewed, and $t_0$ should instead be retained as an explicit index in the mixture of Equation~\ref{eq:summary_equation_final}.
In this regime, the marginalization over $t_0$ also cannot be replaced by maximization, and the considered window size matters, since the look-elsewhere effect can dilute the amplitude.

\section{Information loss due to using folded data}
\label{app:folded_data_loss}
Kepler noise spectrum changes between different 90-day quarters \citep[][]{kepler_data_processing_handbook}, and our pipeline estimates the whitening filter on a per-quarter basis. Thus, transits of the same target may be whitened with different filters and have different expected transit models. 
The correct whitened data model (Equation~\ref{eq:data_model}) should include the full ephemeris with proper whitened model $\mathbf{h}_j$ for each transit $j$. This would increase the computational complexity, and we wanted to avoid it by working with pre-folded data and using the expected average whitened transit model $\mathbf{h}$. In this Appendix, we quantify the information loss associated with this approximation. 

\begin{figure*}[ht]
    \centering
    \includegraphics[width=0.5\textwidth]{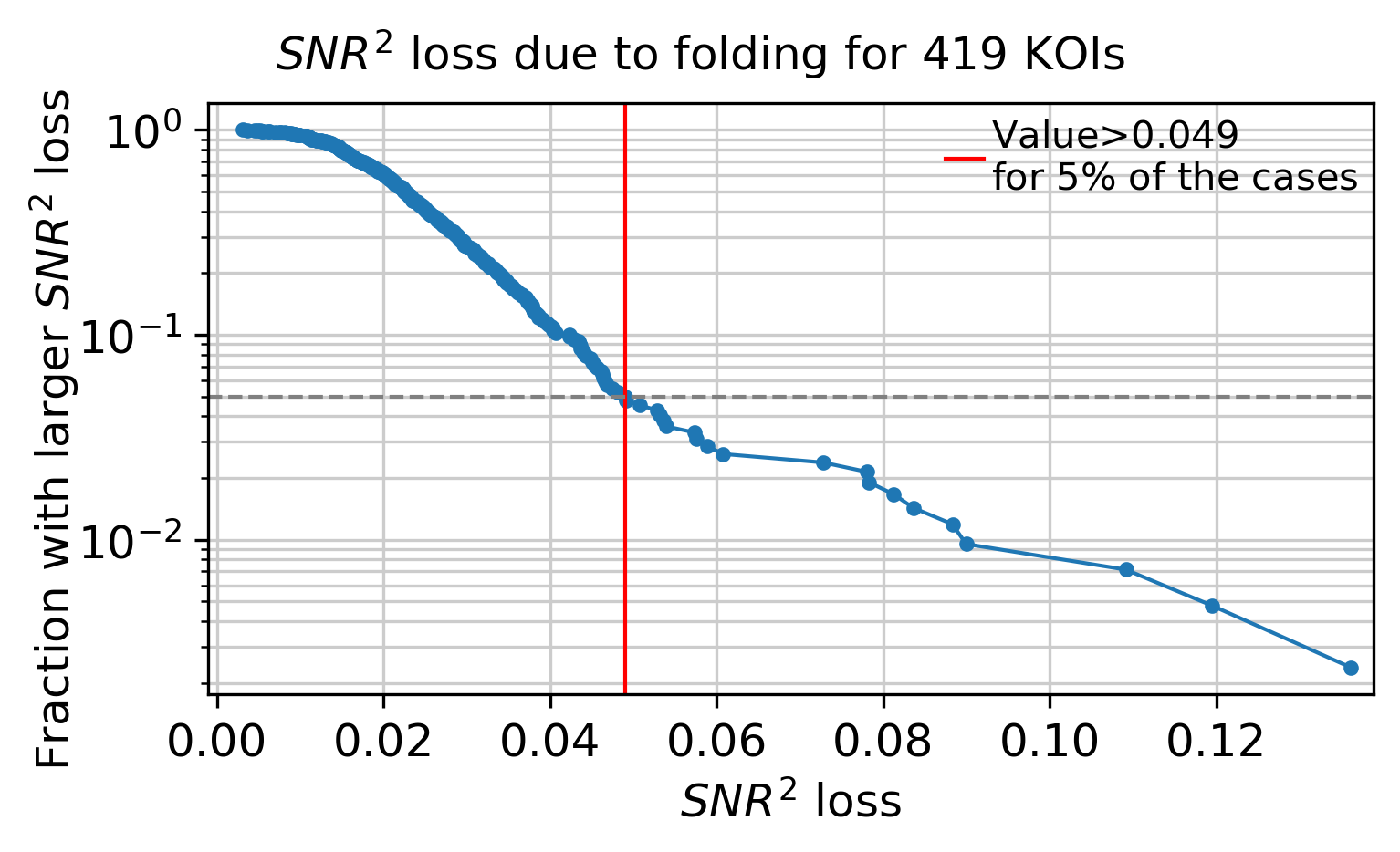}
    \caption{Cumulative distribution of the SNR$^2$ loss of the KOIs from Section~\ref{sec:comparison} due to using folded flux and fitting it with one whitened transit model.}
    \label{fig:folding_error}
\end{figure*}

If a full ephemeris with a varying whitening filter is used, the exact $\lambda$ estimator in Equation~\ref{eq:factorized_likelihood_lambda_hat} is
\begin{align}
    \hat{\lambda}_{\rm ex}
    =\frac{\sum_j\left<\mathbf{d}_j,\mathbf{h}_j\right>}
    {\sum_j\left<\mathbf{h}_j,\mathbf{h}_j\right>},
    \qquad
    \sigma^2_{\lambda,{\rm ex}}
    =\frac{1}{\sum_j\left<\mathbf{h}_j,\mathbf{h}_j\right>},
    \label{eq:exact_estimator}
\end{align}
where $j$ enumerates over the $N_{\text{tr}}$ transits. 

When computing the likelihood on the folded whitened data $\mathbf{d}\!=\!\frac{1}{N_{\text{tr}}}\sum_j\mathbf{d}_j$, we use the average whitened template, or the template whitened with the averaged whitening filter $\mathbf{h}\!=\!\frac{1}{N_{\text{tr}}}\sum_j\mathbf{h}_j$.

If such template is used, folding introduces no bias in $\lambda$, since both estimators (Equations~\ref{eq:amplitude_estimator}, \ref{eq:exact_estimator}) are linear in the data with $\mathbb{E}[\hat{\lambda}_{\rm ex}]=\mathbb{E}[\hat{\lambda}]=\lambda$.

Using folded data discards information of different weights that need to multiply every transit: every transit inside the folded data is multiplied by the same averaged template. This changes the variance of the $\lambda$ estimator and value of the likelihood. Both changes are defined by the ratio of the Fisher information on $\lambda$,
\begin{align}
    \epsilon^2
    \equiv\frac{\sigma^2_{\lambda,{\rm ex}}}{\sigma^2_{\lambda}}
    =\frac{N_{\text{tr}}\left<\mathbf{h},\mathbf{h}\right>_{1}}
    {\sum_j\left<\mathbf{h}_j,\mathbf{h}_j\right>_{1}}
    \le 1,
    \label{eq:folding_efficiency}
\end{align}
where $\left<\cdot,\cdot\right>_1$ denotes the unit-weight inner product. The loss is quadratic in the quarter-to-quarter variation of the whitened template, so first-order differences between the whitening filters produce only second-order effects.

The likelihood change is defined by the $\text{SNR}^2$ loss, $1-\epsilon^2$.

To verify the significance of this effect for the KOIs that were used in Section~\ref{sec:comparison}, we calculated the expected $\text{SNR}^2$ loss for all of them using Equation~\ref{eq:folding_efficiency}. We compared the templates whitened with the filter corresponding to the quarter of each transit, and with the average whitening filter that our algorithm employs. In Figure~\ref{fig:folding_error}, we present the cumulative distribution of the $\text{SNR}^2$ loss over the KOIs. From it, we see that for our sample, only $5\%$ of sources get $\text{SNR}^2$ loss exceeding $5\%$. This corresponds to $5\%$ of the sources getting inflation in the $\lambda$ standard deviation, $\sigma_{\lambda}/\sigma_{\lambda,{\rm ex}}=\epsilon$, more than 1.025. Since this effect is not significant compared to the measured posterior widths, we used folded data for this inference.
Removing these targets from our sample does not change the shapes of the distributions presented in Figure~\ref{fig:normalized_errors_distribution_and_p_vales_koi}.

We note that this approximation is not accurate for all Kepler targets. In some cases, noise spectrum differences between quarters can be significant, particularly when a strong contaminant source is present in the pixel stamp. $\text{SNR}^2$ loss for some targets can exceed 50$\%$. In such cases, our inference should be performed per transit rather than on the folded data, and the per-quarter values should be summarized to get the resulting likelihood and $\lambda$.

\bibliography{kepler_catalog_ref.bib}{}
\bibliographystyle{aasjournal}

\end{document}